\documentclass[manuscript]{acmart}

\setcopyright{none}

\hypersetup{citecolor=blue,linkcolor=blue}
\usepackage{amsmath,amsopn,amssymb,amsthm}
\usepackage{subfigure}
\usepackage{endnotes,microtype,xspace,graphicx,fancyvrb,multirow}
\usepackage{booktabs}
\usepackage{array,underscore,relsize}
\usepackage[T1]{fontenc}
\usepackage{times}
\usepackage{fancyhdr}
\usepackage{enumitem}
\usepackage{listings}
\usepackage{aliascnt}
\usepackage[linesnumbered]{algorithm2e}
\usepackage{semantic}
\usepackage{stmaryrd}
\usepackage{verbatim}

\newcommand{\sys}{\mbox{\textsc{Pequod}}\xspace}

\newcommand{\cc}[1]{\mbox{\smaller[0.5]\texttt{#1}}}

\fvset{fontsize=\scriptsize,xleftmargin=8pt,numbers=left,numbersep=5pt}

\input{fmt}

\def\Snospace~{\S{}}

\newif\ifdraft\drafttrue
\newif\ifnotes\notestrue
\ifdraft\else\notesfalse\fi

\input{glyphtounicode}
\newcolumntype{R}[1]{>{\raggedleft\let\newline\\\arraybackslash\hspace{0pt}}p{#1}}

\newcommand{\squishlist}{
\begin{itemize}[noitemsep,nolistsep]
  \setlength{\itemsep}{-0pt}
}
\newcommand{\squishend}{
  \end{itemize}
}

\usepackage{tikz}

\newcommand{\add}[2]{#1 \union \{ #2 \}}

\newcommand{\assign}{\mathbin{:=}}

\newcommand{\auflia}{\textsc{Auflia}\xspace}

\newcommand{\bigland}{\bigwedge}

\newcommand{\biglor}{\bigvee}

\newcommand{\bigunion}{\bigcup}

\newcommand{\concat}{\mathbin{\cdot}}

\newcommand{\domain}{\mathsf{Dom}}

\newcommand{\elts}[1]{\{ #1 \}}

\newcommand{\entails}{\models}

\newcommand{\false}{\mathsf{False}}

\newcommand{\intersection}{\cap}

\newcommand{\ints}{\mathbb{Z}}

\newcommand{\lia}{\textsc{Lia}\xspace}

\newcommand{\partto}{\hookrightarrow}

\newcommand{\replace}[3]{#1 [ #3 / #2 ]}

\newcommand{\sats}{\vdash}

\newcommand{\setformer}[2]{\{ #1\ |\ #2 \}}

\newcommand{\subs}[2]{#1 [ #2 ]}

\newcommand{\true}{\mathsf{True}}

\newcommand{\union}{\cup}

\newtheorem{defn}{\bf{Definition}}

\newtheorem{ex}{\bf{Example}}

\newtheorem{lemma}{\bf{Lemma}}

\newtheorem{thm}{\bf{Theorem}}

\newcommand{\baseLine}{\textsc{Baseline}\xspace}

\newcommand{\brtargetof}[1]{\mathsf{BrTgt}[#1]}

\newcommand{\chkinductive}{\textsc{ChkInd}\xspace}

\newcommand{\chkinductiveaux}{\textsc{C'}\xspace}

\newcommand{\chooseres}{\textsc{Choose}\xspace}

\newcommand{\discharged}{\mathsf{dis}}

\newcommand{\duality}{\textsc{Duality}\xspace}

\newcommand{\entryof}[1]{\mathsf{Entry}}

\newcommand{\exts}{\mathsf{Ext}}

\newcommand{\finalloc}{\cc{FINAL}}

\newcommand{\formulas}[1]{\mathsf{Forms}[#1]}

\newcommand{\initloc}{\cc{INIT}\xspace}

\newcommand{\instrat}[1]{\mathsf{Instr}[ #1 ]}

\newcommand{\instrs}{\cc{Instrs}}

\newcommand{\instrof}[1]{\mathsf{Instr}[#1]}

\newcommand{\isdis}{\mathsf{IsDis}}

\newcommand{\isind}{\mathsf{HasInd}}

\newcommand{\issat}{\textsc{IsSat}\xspace}

\newcommand{\lang}{\cc{Lang}}

\newcommand{\locinvs}[2]{\mathsf{LocInvs}[ #1, #2 ]}

\newcommand{\locrels}{\mathsf{LocRels}}

\newcommand{\locrelsof}[1]{\mathsf{LocRels}[ #1 ]}

\newcommand{\locs}{\cc{Locs}}

\newcommand{\mergeinvs}{\textsc{Mrg}\xspace}

\newcommand{\mincomplete}{\textsc{Cmpl}\xspace}

\newcommand{\modelof}[1]{m^{#1}}

\newcommand{\noneqpaths}[2]{\mathsf{NoEq}[ #1, #2 ]}

\newcommand{\nonequiv}{\mathsf{NonEq}}

\newcommand{\obligations}{\mathsf{obs}}

\newcommand{\prelocof}[1]{\mathsf{Pre}[ #1 ]}

\newcommand{\params}{\cc{Params}\xspace}

\newcommand{\paths}[1]{\mathsf{Paths}[ #1 ]}

\newcommand{\pathpairinvs}[2]{\mathsf{PathInvs}[ #1, #2 ]}

\newcommand{\pathpairrels}[2]{\mathsf{PathRels}[ #1, #2 ]}

\newcommand{\postctr}[2]{\mathsf{PostCtr}[ #1, #2 ]}

\newcommand{\prefixesof}[1]{\mathsf{Prefixes}[ #1 ]}

\newcommand{\progseqs}{\cc{LblInstrs}}

\newcommand{\refine}{\textsc{PathInvs}\xspace}

\newcommand{\remainctr}[2]{\mathsf{RemainCtr}[ #1, #2 ]}

\newcommand{\remelt}{\textsc{Rem}\xspace}

\newcommand{\retvar}{\cc{ret}}

\newcommand{\runsof}[1]{\mathsf{Runs}[#1]}

\newcommand{\solveaux}{\textsc{Peq'}\xspace}

\newcommand{\solveitp}{\textsc{Itp}\xspace}

\newcommand{\stores}{\mathsf{Stores}}

\newcommand{\subrange}{\mathsf{Subrange}}

\newcommand{\symrelof}[1]{\mathsf{Sem}[ #1 ] }

\newcommand{\symrels}{\mathsf{SymRels}}

\newcommand{\tailof}[1]{\mathsf{tl}[#1]}

\newcommand{\transrelof}[1]{\rho[#1]}

\newcommand{\values}{\mathsf{Vals}}

\newcommand{\vars}{\cc{Vars}}

\newcommand{\vocab}{\mathcal{V}}

\gdef\therev{2b72454}
\gdef\thedate{2017-05-08 16:38:36 -0400}

\begin{document}

\title{Completely Automated Equivalence Proofs}
\author{Qi Zhou}
\affiliation{%
  \institution{Georgia Institute of Technology}
  }
\author{David Heath}
\affiliation{%
  \institution{Georgia Institute of Technology}
}
\author{William Harris}
\affiliation{%
  \institution{Georgia Institute of Technology}
}
\ifdefined\DRAFT
 \pagestyle{fancyplain}
 \lhead{Rev.~\therev}
 \rhead{\thedate}
 \cfoot{\thepage\ of \pageref{LastPage}}
\fi


\date{}

\begin{abstract}
  Verifying partial (i.e., termination-insensitive) equivalence of
  programs has significant practical applications in software
  development and education.
  Conventional equivalence verifiers typically rely on a combination
  of given relational summaries and suggested synchronization points;
  such information can be extremely difficult for programmers without
  a background in formal methods to provide for pairs of programs with
  dissimilar logic.

  In this work, we propose a completely automated verifier for
  determining partial equivalence, named \sys.
  \sys automatically synthesizes expressive proofs of equivalence
  conventionally only achievable via careful, manual constructions of
  product programs
  To do so, \sys syntheses relational proofs for selected pairs of
  program paths and combines the per-path relational proofs to
  synthesize relational program invariants.
  To evaluate \sys, we implemented it as a tool that targets Java
  Virtual Machine bytecode and applied it to verify the equivalence of
  hundreds of pairs of solutions submitted by students for problems
  hosted on popular online coding platforms, most of which could not
  be verified by existing techniques.
\end{abstract}

\maketitle

\section{Introduction}
\label{sec:introduction}
In many practical contexts, determining if two programs are
functionally equivalent is a critical problem.
Prominent instances of this problem include determining (1) if a given
program written in a high-level source language is equivalent to a
given (typically optimized) program that executes on a target machine
architecture~\cite{leroy06,pnueli98},
(2) if consecutive versions of a program module preserve critical
program behavior,
(3) if one given program is an obfuscation of the other, or
(4) if a program provided by a student or hiring candidate in response
to a challenge problem is equivalent to a trusted reference solution.
Checking student solutions, in particular, is perhaps more critical
than ever before, given increasing enrollments in computer science
courses and the rapid development of online programming
courses~\cite{singh13}.

While verifying even only termination-insensitive (i.e.,
\emph{partial}) equivalence has been the subject of a significant body
of work, many previous techniques are either
intended to be applied to verify equivalence of programs generated
from particular transformations~\cite{leroy06,necula00,pnueli98},
or can only be applied to programs that use restricted control
structures (e.g., are loop-free~\cite{lahiri12}) or data operations
(e.g., only linear arithmetic on scalar data, without operations on
dynamically-allocated memory~\cite{partush13,verdoolaege11}).
Other approaches only generate proofs for a bounded number of control
paths~\cite{person08,ramos11} or inputs~\cite{singh13}.

One strategy that can potentially be followed to prove the equivalence
of many programs is to reduce the problem of verifying equivalence of
programs $P_0$ and $P_1$ to synthesizing a \emph{product program} that
soundly models all steps of $P_0$ and $P_1$, accompanied by inductive
invariants of the product program that imply the equivalence of $P_0$
and $P_1$.
Unfortunately, current approaches that follow such a strategy either
only attempt to synthesize product programs in a class that is too
restricted to prove equivalence of many practical programs, such as
the class of sequential
compositions~\cite{barthe04,felsing14,terauchi05}, or require
additional information about a target product program to be provided
manually~\cite{benton04,barthe11,godlin09,hawblitzel13,sousa16}.

In this paper, we present a novel verifier for partial equivalence,
named \sys, which is not subject to the limitations given above.
I.e., \sys can be applied to pairs of programs with arbitrary control
structure and that use arbitrary data operations, and can potentially
synthesize proofs ranging over a class of product programs that is
much more expressive than those that have been synthesized by previous
automatic verifiers.

The key challenge addressed by \sys is, given programs $P_0$ and
$P_1$, to synthesize both a product program of $P_0$ and $P_1$ and
suitable inductive invariants automatically.
Previous approaches either require the structure of a product program
to be provided manually, or that first attempt to guess the structure
of a product program using heuristics, and then synthesize invariants
for the product program by adapting techniques used by automatic
verifiers of safety properties.
Unfortunately, it is difficult to communicate the requirements of a
product program to users without experience in program analysis (such
as novice programmers).
Proposed heuristics can only be applied in practice to programs have
syntactic similarities that typically only hold for multiple versions
of the same program.
However, it is difficult to develop heuristics that can be applied to
programs that have been developed by independent developers, such as a
solutions submitted by independent groups of students.

\sys addresses this key challenge by synthesizing both the product
program and its inductive invariants \emph{simultaneously}.
In particular, \sys selectively enumerates control paths of $P_0$
paired with those of $P_1$.
For each enumerated pair of control paths $p_0$ and $p_1$, \sys first
determines if some runs of the paths from equivalent inputs result in
non-equivalent outputs, in which case it determines that $P_0$ and
$P_1$ are not equivalent.
Otherwise, \sys efficiently synthesizes a proof that each run of $p_0$
and each run of $p_1$ from equivalent inputs result in equivalent
outputs.
\sys combines proofs synthesized for multiple pairs of paths, and then
attempts to extract from them a product program and its inductive
invariants using a novel symbolic search algorithm.
An extensive body of previous work has developed automatic verifiers
that synthesize inductive invariants of a single program from
invariants of program paths in order to prove that a program satisfies
a given safety
property~\cite{ball02,henzinger02,henzinger04,mcmillan06}.
The contribution of the proposed work is to adapt such a strategy to
simultaneously synthesize a product program and its invariants in
order to prove that given programs are equivalent.

We have implemented a prototype of \sys that verifies the partial
equivalence of programs given in Java Virtual Machine (JVM) bytecode
and have applied \sys to verify the partial equivalence of $369$ pairs
of solutions to challenge problems hosted on online coding
platforms~\cite{codechef,leetcode}.
Implementations of previous automated equivalence verifiers could
verify only one of pairs of programs that we found.

The rest of this paper is organized as follows.
\autoref{sec:overview} provides an informal overview of our
approach, \sys, by example.
\autoref{sec:background} reviews the technical foundations for our
work, and \autoref{sec:approach} presents \sys in detail.
\autoref{sec:evaluation} presents an empirical evaluation of \sys.
\autoref{sec:related-work} compares \sys to related work on
equivalence verification, and %
\autoref{sec:conclusion} concludes.


\section{Overview}
\label{sec:overview}
In this section, we illustrate \sys by example.
In \autoref{sec:ex-programs}, we present as a running example a pair
of programs that were submitted independently as solutions to an
online coding problem.
In \autoref{sec:ex-rel-invs}, we give a proof that the two solutions
are partially equivalent, expressed as relational invariants over
pairs of control locations.
In \autoref{sec:ex-verify}, we illustrate how \sys synthesizes the
proof automatically.

\subsection{Climbing Stairs: a coding challenge problem}
\label{sec:ex-programs}

\begin{figure}[t]
  \begin{minipage}{0.5\linewidth}
    \input{climb-stairs-0.java}    
  \end{minipage}
  \begin{minipage}{0.4\linewidth}
    \input{climb-stairs-1.java}    
  \end{minipage}
  \caption{\cc{climbStairs0} and \cc{climbStairs1}: two solutions
    provided for the Climbing Stairs Problem hosted on the LeetCode
    coding platform.
  }
  \label{fig:running-ex-code}
\end{figure}
\autoref{fig:running-ex-code} contains the pseudocode for two
solutions to the Climbing Stairs Problem hosted on the coding platform
LeetCode~\cite{leetcode}.
The Climbing Stairs Problem is to take an integer $n$ and return the
number of distinct ways to climb $n$ steps, where steps can be climbed
one or two at a time.  If $n \leq 1$, then the solution is one.

\cc{climbStairs0} and \cc{climbStairs1} are two correct solutions to
the problem, submitted by independent programmers.
\cc{climbStairs0} first checks if its argument \cc{n} is less than or
equal to $1$, and if so, immediately returns $1$ (line 5).
Otherwise, \cc{climbStairs0} executes a loop with counter \cc{i}
incremented from $2$ to $n$ (lines 9---14).
The loop maintains the invariant that at each step, \cc{sum} stores
the number of sequences in which to climb \cc{i} stairs, \cc{cur}
stores the number of sequences in which to climb $\cc{i} - 1$ stairs,
and \cc{prev} stores the number of sequences in which to climb $\cc{i}
- 2$ stairs.
In each step through the loop, \cc{climbStairs0} copies the value in
\cc{sum} to \cc{cur} (line 11), increments the value in \cc{sum} by
\cc{prev} (line 12), and copies the value in \cc{cur} to \cc{prev}
(line 13).
\cc{climbStairs0} iterates until $\cc{i} \geq n$ and then returns the
value stored in \cc{cur} (lines 15).

\cc{climbStairs1} is similar to \cc{climbStairs0}, but maintains the
invariant that the variable \cc{count2} stores the number of sequences
in which to climb $\cc{i} - 1$ stairs and \cc{count1} stores the
number of sequences in which to climb $\cc{i} - 2$ stairs.
While \cc{count1} and \cc{count2} are used in \cc{climbStairs1}
similarly to how \cc{cur} and \cc{sum} are used in \cc{climbStairs0},
they are initialized to distinct values to establish
\cc{ClimbStair1}'s loop invariant (lines 7---8).
Given the same input, \cc{climbStairs1} performs one more iteration of
its loop than \cc{climbStairs0}.

\subsection{Equivalence of \cc{climbStairs0} and \cc{climbStairs1}}
\label{sec:ex-rel-invs}
\cc{climbStairs0} and \cc{climbStairs1}, when given equal inputs on
which they terminate, exit in states with return equal values;
i.e., the programs are \emph{partially equivalent}.

\sys, given programs $P_0$ and $P_1$ attempts to determine if they are
partially equivalent by synthesizing a \emph{product program} of $P_0$
and $P_1$, denoted $P'$, accompanied by suitable inductive
invariants~\cite{barthe11,barthe13,barthe16}.
A product program of $P_0$ and $P_1$ is a program in which each
location is a pair of a location of $P_0$ and a location of $P_1$ and
each state is a pair of a state of $P_0$ and a state of $P_1$.
In each step of execution, the product program chooses a stepping
component program---either $P_0$ or $P_1$---based on its state, and
then non-deterministically chooses an instruction of the chosen
component program on which to step.
Thus, there are potentially infinitely many product programs of fixed
programs $P_0$ and $P_1$.
Each product program has the same state space, but in each step,
chooses the stepping program based on a different predicate on its
current state.

The equivalence of $P_0$ and $P_1$ is certified by inductive
invariants of $P'$ \textbf{(1)} the invariant at the pair of initial
locations of $P_0$ and $P_1$ is supported by the assumption that the
components of state corresponding to $P_0$ and $P_1$ have equivalent
arguments; %
\textbf{(2)} the invariant at the pair of final locations of $P_0$ and
$P_1$ supports the conclusion that the components of state
corresponding to $P_0$ and $P_1$ have equivalent return values.
Such invariants are represented as a map from each pair of locations
to a formula over a vocabulary consisting of the variables of the two
programs.

The programs \cc{climbStairs0} and \cc{climbStairs1} have proofs of
equivalence, each structured as a product program accompanied by such
inductive invariants.
For one such proof, the product program $\cc{climbStairs}'$, at each
pair of identical line numbers, chooses to step on \cc{climbStairs0},
and at all other pairs of locations, chooses to step on
\cc{climbStairs1}.
We will now describe a proof of equivalence of \cc{climbStairs0} and
\cc{climbStairs1} as inductive invariants of the fixed product program
$\cc{climbStairs}'$.
However, a key feature of \sys is that it does not require a fixed
product program to be given manually or as the result of heuristics.
Instead, \sys synthesizes both a product program and its invariants
simultaneously.
Such a technique is essential for automatically verifying the
equivalence of programs that, unlike the relatively simple examples of
\cc{climbStairs0} and \cc{climbStairs1}, have dissimilar control
structure or data variables.

Inductive invariants of $\cc{climbStairs}'$ can be represented as a
map from pairs of control locations to symbolic relations.
We give symbolic relations over key pairs of locations as formulas
over a logical vocabulary consisting of variables that occur in
\cc{climbStairs0} and \cc{climbStairs1}, denoted with subscripts $0$
and $1$.
In this paper, we only consider symbolic relations defined over
constraints in linear arithmetic, because this is sufficient to
axiomatize the semantics of the simple programs that we describe.
Our implementation of \sys for JVM bytecode synthesizes invariants in
a more expressive logic that can describe states with
dynamically-allocated objects and arrays, namely the combination of
the theories of linear arithmetic and arrays.

The relational invariant over lines \cc{2} and \cc{2}, denoted
$I(\cc{2}, \cc{2})$, establishes that the components of the state of
$\cc{climbStairs}'$ for \cc{climbStairs0} and \cc{climbStairs1} have
equal arguments.
I.e., $I(\cc{2}, \cc{2})$ is
\begin{equation*}
  \cc{n}_0 = \cc{n}_1 \label{eqn:init-ctr}
\end{equation*}

The relational invariant for line \cc{5} in \cc{climbStairs0} and
\cc{5} in \cc{climbStairs1}, denoted $I(\cc{5}, \cc{5})$, establishes
that any pair of states in \cc{climbStairs0} and \cc{climbStairs1} at
such locations will result in states with equivalent return values.
I.e., $I(\cc{5}, \cc{5})$ is
\begin{equation*}
  \cc{result}_0 = \cc{result}_1
\end{equation*}

The relational invariant for line \cc{9} of \cc{climbStairs0} and line
\cc{9} of \cc{climbStairs1}, denoted $I(\cc{9}, \cc{9})$, establishes
that for each run of $\cc{climbStairs}'$, %
\textbf{(1)} the value of \cc{i} in \cc{climbStairs0} is one greater
than the value of \cc{i} in \cc{climbStairs1}, %
\textbf{(2)} the value of \cc{sum} in \cc{climbStairs0} is equal to
the value of \cc{count2} in \cc{climbStairs1}, and %
\textbf{(3)} the values of \cc{n} in \cc{climbStairs0} and
\cc{climbStairs1} are equal.
I.e., $I(\cc{9}, \cc{9})$ is
\[ \cc{i}_0 = \cc{i}_1 + 1 \land \cc{sum}_0 = \cc{count2}_1 \land %
   \cc{n}_0 = \cc{n}_1
\]

The relational invariant for line \cc{15} of \cc{climbStairs0} and
line \cc{15} of \cc{climbStairs}, denoted $I(\cc{15}, \cc{15})$,
establishes that the components of the state of $\cc{climbStairs}'$
for \cc{climbStairs0} and \cc{climbStairs1} have equal values in their
return variables.
I.e., $I(\cc{15}, \cc{15})$ is
\begin{equation*}
  \cc{result}_0 = \cc{result}_1
\end{equation*}

The symbolic relations for the pairs of locations given above define
inductive invariants for $\cc{climbStairs}'$ that are supported by the
assumption that \cc{climbStairs0} and \cc{climbStairs1} execute from
states with equal arguments, and that support the assertion that if
\cc{climbStairs0} and \cc{climbStairs1} terminate, they have equal
return values.
Thus, the invariants are proof that \cc{climbStairs0} and
\cc{climbStairs1} are partially equivalent.

\subsection{Synthesizing a product program and its invariants}
\label{sec:ex-verify}
\begin{figure}[t]
  \centering
  \includegraphics[width=\linewidth]{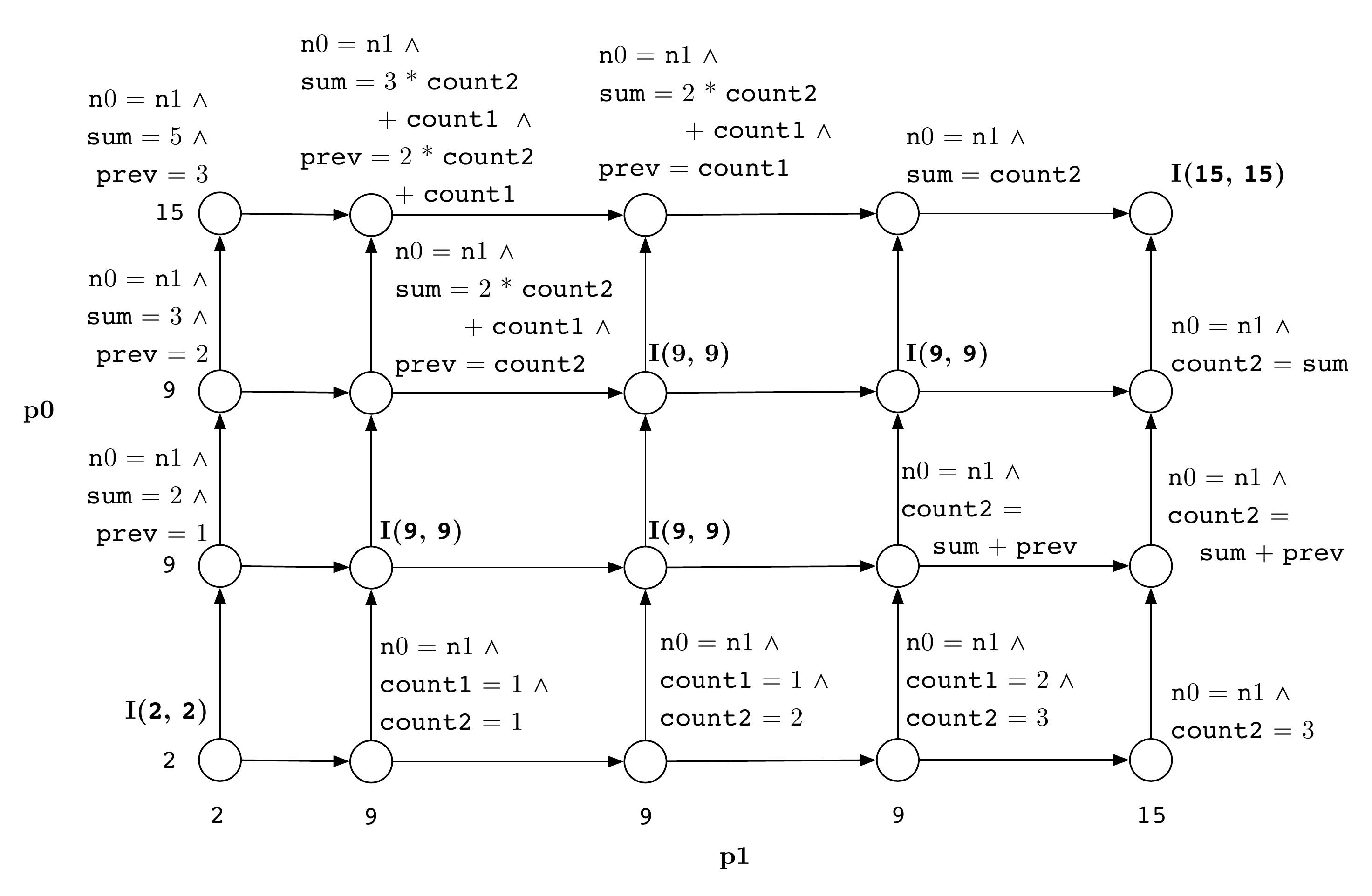}
  \caption{Invariants for pairs of prefixes of a control
    path $p_0$ of \cc{climbStairs0} paired with a path $p_1$ of
    \cc{climbStairs1}.
    Each node $n$ represents a pair of subpaths of $p_0$ and $p_1$.
    The sequence of control locations from the bottom up to row of $n$
    contain its path in $p_0$;
    the sequence of control locations from the left to the
    column of $n$ contains its path in $p_1$.
    $n$ is annotated with a relational invariant over all runs of its
    pair of paths.
  }
  \label{fig:path-pair-invs}
\end{figure}
\sys, given programs $P_0$ and $P_1$, attempts to synthesize a product
program of $P_0$ and $P_1$ accompanied by inductive invariants by
iteratively maintaining invariants of sets of pairs of $P_0$ and
$P_1$'s \emph{paths}.
If the invariants $I$ are defined for a path $p_0$ of $P_0$ and path
$p_1$ of $P_1$, then $I$ maps $p_0$ and $p_1$ to a symbolic relation
between all pairs of states reached after $P_0$ executes $p_0$ and
$P_1$ executes $p_1$ from states with equal arguments.

For example, \autoref{fig:path-pair-invs} depicts path-pair invariants
for all pairs of prefixes of a complete path $p_0$ of
\cc{climbStairs0} and a complete path $p_1$ of \cc{climbStairs1}.
$p_0$ is the control path of \cc{climbStairs0} that executes the loop
in lines 9---14 once, and $p_1$ is the control path of
\cc{climbStairs1} that executes the loop in lines 9---14 twice.
I.e., $p_0$ and $p_1$ are the paths executed by their programs on
input $\cc{n} = 3$.
The relational invariants for path pairs $([ \cc{2} ], [ \cc{2} ])$,
$([ \cc{2}, \cc{9} ], [ \cc{2}, \cc{9} ])$, $([ \cc{2}, \cc{9} ], [
\cc{2}, \cc{9}, \cc{9} ])$, $([ \cc{2}, \cc{9}, \cc{9} ], [ \cc{2},
\cc{9}, \cc{9} ])$, $([ \cc{2}, \cc{9}, \cc{9} ], [ \cc{2}, \cc{9},
\cc{9}, \cc{9}])$, and $([ \cc{2}, \cc{9}, \cc{9}, \cc{15} ], [
\cc{2}, \cc{9}, \cc{9}, \cc{9}, \cc{15}])$ are the entries in the
location-pair invariants $I$ for \cc{climbStairs0} and
\cc{climbStairs1} given in \autoref{sec:ex-rel-invs}.
The invariants for all other pairs of prefixes of $p_0$ and $p_1$ are
given explicitly in \autoref{fig:path-pair-invs}.

In each of \sys's iterations, it determines if the maintained
path-pair invariants $I_M$ define inductive invariants of some product
program of $P_0$ and $P_1$.
In particular, for path-pair invariants $I_p$, if the map
$\locrelsof{I_p}$ from each pair of locations $L_0$ and $L_1$ to the
disjunction of invariants in $I_p$ for all pairs of paths ending with
$L_0$ and $L_1$ are inductive invariants of some product program of
$P_0$ and $P_1$, then $I_p$ are \emph{inductive path-pair invariants}.
If \sys finds a subset of bindings of $I_M$ (i.e., some
\emph{restriction} of $I_M$) that is inductive, then \sys determines
that $P_0$ and $P_1$ are equivalent.
If not, \sys selects a path $p_0$ of $P_0$ and $p_1$ of $P_1$ on which
its maintained path-pair invariants are undefined, attempts to
synthesize invariants for $p_0$ and $p_1$, and if it finds such
invariants, merges them with the maintained set of path-pair
invariants to complete its current iteration.
\sys's algorithm is described in detail in \autoref{sec:solve}.

E.g., \sys, given \cc{climbStairs0} and \cc{climbStairs1}, synthesizes
inductive path-pair invariants for the programs over the following
steps.
\sys chooses as initial path-pair invariants the empty map.
\sys then determines that $\emptyset$ does not define inductive
path-pair invariants, using a procedure discussed informally below and
given in detail in \autoref{sec:chk-inductive}.
As a result, \sys chooses $p_0$ as a path of \cc{climbStairs0} and
$p_1$ as a path of \cc{climbStairs1} that have no path-pair invariant
in $\emptyset$.
\sys then attempts to synthesize path-pair invariants for $p_0$ and
$p_1$.

\sys could be adapted to use different path-selection algorithms,
causing it to choose different pairs of complete paths.
We will consider a scenario in which \sys chooses $p_0$ and $p_1$ in
particular, because those paths most clearly illustrate the operation
of \sys.

\paragraph{Proving equivalence of pairs of paths}
After \sys selects a pair of paths $p_0$ and $p_1$ that are undefined
in its maintained set of path-pair invariants, \sys determines if
$p_0$ and $p_1$ are equivalent, and synthesizes path-pair invariants
for $p_0$ and $p_1$ by issuing repeated queries to an interpolating
theorem prover.
\sys synthesizes path-pair invariants for $p_0$ and $p_1$ that contain
path-pair invariants for each prefix of $p_0$ paired with each prefix
of $p_1$.
Each path-pair invariant for a pair of path prefixes is synthesized
from a \emph{logical interpolant}, generated by a query to an
interpolating theorem prover.
The definition of interpolants is reviewed in \autoref{sec:logic},
\autoref{defn:itps};
the reduction from synthesizing path-pair invariants to finding
interpolants is given in \autoref{sec:find-path-invs}.

E.g., to verify that path $p_0$ of \cc{climbStairs0} and path $p_1$
\cc{climbStairs1} are equivalent, \sys synthesizes invariants for each
pair of a prefix of $p_0$ with a prefix of $p_1$.
One such collection of invariants over pairs of prefixes is depicted
in \autoref{fig:path-pair-invs}.

\paragraph{From path-pair invariants to a product program and its invariants}
After \sys extends its maintained path-pair invariants to include
path-pair invariants for chosen paths $p_0$ and $p_1$, it inspects the
extended path-pair invariants to determine if some restriction are
inductive, using a novel algorithm described in
\autoref{sec:chk-inductive}.

E.g., for paths $p_0$ of \cc{climbStairs0} and $p_1$ of
\cc{climbStairs1}, some restriction of the path pair invariants $I_p$
defines inductive path-pair invariants for a sub-program of
\cc{climbStairs0} paired with a subprogram of \cc{climbStairs1}.
In particular, let $\cc{climbStairs0}_{\cc{else}}$ be
\cc{climbStairs0}, transformed so that the \cc{then} branch is
replaced with an instruction that halts without returning, and
similarly for $\cc{climbStairs1}_{\cc{else}}$ and \cc{climbStairs1}.
Let $\cc{climbStars}_{\cc{else}}'$ be the product program of
$\cc{climbStairs0}_{\cc{else}}$ and $\cc{climbStairs1}_{\cc{else}}$,
defined similarly to $\cc{climbStairs}'$ for \cc{climbStairs0} and
\cc{climbStairs1}.
Let $I_{\cc{else}}$ be the restriction of $I_p$ to the invariants for
the pairs of paths $([ \cc{2}, \cc{9} ], [ \cc{2}, \cc{9}, \cc{9}
])$, %
$([ \cc{2}, \cc{9}, \cc{9} ], [ \cc{2}, \cc{9}, \cc{9} ])$, %
$([ \cc{2}, \cc{9}, \cc{9} ], [ \cc{2}, \cc{9}, \cc{9}, \cc{9} ])$, %
$([ \cc{2}, \cc{9}, \cc{9} ], [ \cc{2}, \cc{9}, \cc{9}, \cc{9} ])$,
and %
$([ \cc{2}, \cc{9}, \cc{9}, \cc{15} ], [ \cc{2}, \cc{9}, \cc{9},
\cc{9}, \cc{15} ])$.
Then $\locrelsof{I_{\cc{else}}}$ are inductive invariants of
$\cc{climbStairs}_{\cc{else}}'$, and thus prove the equivalence of
$\cc{climbStairs0}_{\cc{else}}$ and $\cc{climbStairs1}_{\cc{else}}$.
\sys, given $\cc{climbStairs0}_{\cc{else}}$ and
$\cc{climbStairs1}_{\cc{else}}$ would automatically synthesize from
$I_p$ both $\cc{climbStairs}_{\cc{else}}'$ and its inductive
invariants $\locrelsof{I_{\cc{else}}}$ as a proof of equivalence.

However, $I_{\cc{else}}$ are not inductive invariants for product
program $\cc{climbStairs}'$, because they map the pair of paths $([
\cc{2} ], [ \cc{2} ])$ to $\cc{n}_0 = \cc{n}_0$, and are not defined
for any pair of paths that contain line \cc{5} in \cc{climbStairs0}
and line \cc{5} in \cc{climbStairs1}.
Thus, $\locrelsof{I_{\cc{else}}}(\cc{2}, \cc{2}) = \cc{n}_0 =
\cc{n}_1$ and $\locrelsof{I_{\cc{else}}}(\cc{15}, \cc{15}) = \false$;
as a result, $\locrelsof{I_{\cc{else}}}$ are not inductive invariants
of $\cc{climbStairs}'$.

\sys, given \cc{climbStairs0} and \cc{climbStairs1}, determines that
in fact no restriction of $I_p$ are inductive path-pair invariants.
\sys continues to determine the equivalence of \cc{climbStairs0} and
\cc{climbStairs1} by choosing a pair of paths $p_0'$ of
\cc{climbStairs0} and $p_1'$ \cc{climbStairs1} that each reach line
\cc{5}.
\sys then synthesizes path-pair invariants $I_p'$ for $p_0'$ and
$p_1'$.
\sys then uses $I_p$ and $I_p'$ to synthesize path-pair invariants
$I_p''$ for both $(p_0, p_1)$ and $(p_0', p_1')$, determines that some
restriction of $I_p''$ are inductive invariants for the product
program $\cc{climbStairs}'$, and thus determines that
\cc{climbStairs0} is equivalent to \cc{climbStairs1}.


\section{Background}
\label{sec:background}
In this section, we review technical concepts on which our approach is
based.
In \autoref{sec:language}, we define a target language of imperative
programs.
In \autoref{sec:logic}, we review concepts from formal logic.

\subsection{Target language}
\label{sec:language}
In this section, we define the structure (\autoref{sec:structure}) and
semantics (\autoref{sec:semantics}) of a language of imperative
programs.

\subsubsection{Program structure}
\label{sec:structure}
A program is a set of instructions that bind the results of
computations to variables.
Let $\locs$ be a space of \emph{control locations} that contain a
distinguished \emph{initial location} \initloc and \emph{final
  location} \finalloc.
Let $\vars$ be a space of program variables, which contains
\emph{parameter} variables \params and a \emph{return} variable
$\retvar$.
Let $\instrs$ be a space of \emph{program instructions}.

A program instruction tests and updates variables and then
transfers a current control location to a target control location.
A pre-location, instruction, and branch-target-location is a
\emph{labeled instruction};
i.e., the labeled instructions are $\progseqs = \locs \times \instrs
\times \locs$.
For each labeled instruction $\cc{i} \in \progseqs$, the pre-location,
instruction, and post-location of $\cc{i}$ are denoted
$\prelocof{\cc{i}}$, $\instrof{\cc{i}}$, and $\brtargetof{\cc{i}}$,
respectively.

A program $\cc{P}$ is a set of labeled instructions such that for all
$\cc{i}_0, \cc{i}_1 \in \cc{P}$, if $\prelocof{\cc{i}_0} = \prelocof{
  \cc{i}_1 }$ and $\brtargetof{ \cc{i}_0 } = \brtargetof{ \cc{i}_1 }$,
then $\cc{i}_0 = \cc{i}_1$.
We denote each $\cc{i} \in \cc{P}$ alternatively as
$\instrat{\cc{P}}(\prelocof{ \cc{i} }, \brtargetof{ \cc{i} })$.
There is no labeled instruction $\cc{i} \in \cc{P}$ for which
$\prelocof{i} = \finalloc$.
The space of programs is denoted \lang.
For the remainder of this section, let $\cc{P} \in \lang$ denote a
fixed, arbitrary program.

\subsubsection{Program semantics}
\label{sec:semantics}
A run of \cc{P} is a sequence of states generated by a sequence of
labeled instructions in which adjacent instructions have matching
target and pre locations.
Let the space of program values be the space of integers;
i.e., the space of values is $\values = \ints$.
An evaluation of all variables in $\vars$ is a store;
i.e., the space of stores is $\stores = \vars \to \values$.
The practical implementation of \sys verifies partial equivalence of
programs that operate on objects and arrays combined with integers.
In this paper, we primarily consider programs that operate over only
integers, and describe how our implementation handles practical
language features in \autoref{sec:practice}.

For each $\cc{i} \in \instrs$, there is a transition relation
$\transrelof{\cc{i}} \subseteq \stores \times \stores$.
For each $\cc{i} \in \progseqs$, the transition relation of the
instruction in $\cc{i}$ is denoted $\transrelof{\cc{i}} =
\transrelof{\instrof{\cc{i}}}$.
The transition relation of an instruction need not be total: thus,
labeled instructions can implement control branches using instructions
that act as \cc{assume} instructions.

A path of \cc{P} is a sequence of control locations that are in
adjacent labeled instructions of \cc{P}.
\begin{defn}
  \label{defn:path}
  Let $\cc{i}_0, \ldots, \cc{i}_n \in \cc{P}$ be such that %
  \textbf{(1)} $\prelocof{\cc{i}_0} = \initloc$ and %
  \textbf{(2)} for each $0 \leq j < n$, $\brtargetof{\cc{i}_j} =
  \prelocof{\cc{i}_{j + 1}}$.
  Then $[ \prelocof{\cc{i}_0}, \ldots, \prelocof{\cc{i}_{n}},
  \brtargetof{\cc{i}_n} ]$ is a \emph{path} of \cc{P}.
\end{defn}
The space of paths of \cc{P} is denoted $\paths{ \cc{P} }$.
The last location in $p$ is denoted $\tailof{p}$.
If $\tailof{p} = \finalloc$, then $p$ is a \emph{complete} path.
For each $p \in \paths{\cc{P}}$, the non-empty prefixes of $p$ are
denoted $\prefixesof{p}$.
For all $p, p' \in \paths{ \cc{P} }$, the set of paths $p'' \in
\paths{ \cc{P} }$ such that $p$ is a prefix of $p''$ and $p''$ is a
prefix of $p'$ is denoted $\subrange(p, p')$.

A run of a program \cc{P} is a path and a sequence of stores $\Sigma$
of equal length, such that adjacent stores in $\Sigma$ satisfy
transition relations of instructions at their corresponding locations
in $p$.
\begin{defn}
  \label{defn:runs}
  Let $\Sigma = \sigma_0, \ldots, \sigma_{n - 1} \in \stores$ and %
  $\cc{L}_0, \ldots, \cc{L}_{n - 1} \in \paths{ \cc{P} }$ be such that
  for each $0 \leq i < n - 1$, $(\sigma_i, \sigma_{i + 1}) \in
  \transrelof{ \instrat{ \cc{P} }(\cc{L}_i, \cc{L}_{i + 1})}$.
  Then $(p, \Sigma)$ is a \emph{run} of \cc{P}.
\end{defn}
The space of runs of \cc{P} is denoted $\runsof{ \cc{P} }$.
For each path $p \in \paths{ \cc{P} }$, the runs $r \in \runsof{
  \cc{P} }$ such that $p$ is the path of $r$ are the runs of $p$.

$\cc{P}_0, \cc{P}_1 \in \lang$ are partially equivalent if all
complete runs of $\cc{P}_0$ and $\cc{P}_1$ that begin from stores in
which parameters have equal values end in stores in which the return
variables store equal values.
\begin{defn}
  \label{defn:part-equiv}
  For all $\cc{P}_0, \cc{P}_1 \in \lang$ and %
  complete $p_0 \in \paths{ \cc{P}_0 }$ and $p_1 \in \paths{ \cc{P}_1
  }$, %
  if for all $\sigma^0_0, \ldots, \sigma^0_m,$ $\sigma^1_0, \ldots,
  \sigma^1_n \in \stores$ such that $(p_0, [ \sigma^0_0, \ldots,
  \sigma^0_m ]) \in \runsof{ \cc{P}_0 }$ and $(p_1, [ \sigma^1_0,
  \ldots, \sigma^1_n ]) \in \runsof{ \cc{P}_1 }$ and
  $\sigma^0_0(\params) = \sigma^1_0(\params)$, %
  it holds that $\sigma^0_m(\retvar) = \sigma^1_n(\retvar)$, %
  then $p_0$ is equivalent to $p_1$ under $\cc{P}_0$ and $\cc{P}_1$,
  denoted $p_0 \equiv p_1$.

  If for all complete $p_0 \in \paths{ \cc{P}_0 }$ and $p_1 \in
  \paths{ \cc{P}_1 }$ it holds that $p_0 \equiv p_1$, then $\cc{P}_0$
  is equivalent to $\cc{P}_1$, denoted $\cc{P}_0 \equiv \cc{P}_1$.
\end{defn}
In order to simplify the presentation of our approach, we have
given a definition of equivalence in terms of equality over identical
parameter and return variables.
However, our approach can be immediately generalized to take as a
specification of equivalence any equivalence relation over input and
final states of two programs.
Because \autoref{defn:part-equiv} defines equivalence in terms of
equal input states and equal resulting output states, it can describe
pairs of programs with different control structures and variables used
for internal computation, such as \cc{climbStairs0} and
\cc{climbStairs1} (introduced in \autoref{sec:ex-programs}).

\subsection{Formal logic}
\label{sec:logic}
\sys uses formal logic to model the semantics of programs and
represent invariants that relate their states.
The quantifier-free fragment of the theory of linear arithmetic is
denoted $\lia$.
For each space of logical variables $X$, the space of $\lia$ formulas
over $X$ is denoted $\formulas{X}$.
For each formula $\varphi \in \formulas{X}$, the set of variables that
occur in $\varphi$ (i.e., the \emph{vocabulary} of $\varphi$) is
denoted $\vocab(\varphi)$.
A \lia \emph{model} $m$ over $X$ is an assignment from each variable
in $X$ to an integer.
The fact that model $m$ \emph{satisfies} a formula $\varphi$ is
denoted $m \sats \varphi$.
For formulas $\varphi_0, \ldots, \varphi_n, \varphi \in \formulas{X}$,
the fact that $\varphi_0, \ldots, \varphi_n$ \emph{entail} $\varphi$
is denoted $\varphi_0, \ldots, \varphi_n \entails \varphi$.

For all vectors of variables $X = [ x_0, \ldots, x_n ]$ and $Y = [
y_0, \ldots, y_n ]$, the \lia formula constraining the equality of
each element in $X$ with its corresponding element in $Y$, i.e., the
formula $\bigland_{0 \leq i \leq n} x_i = y_i$, is denoted $X = Y$.
The repeated replacement of variables $\varphi[ y_0 / x_0 \ldots y_{n
  - 1} / x_{n - 1} ]$ is denoted $\replace{\varphi}{Y}{X}$.
For each formula $\varphi$ defined over free variables $X$,
$\replace{\varphi}{X}{Y}$ is denoted alternatively as $\varphi[Y]$.

Although determining the satisfiability of a \lia formula is
NP-complete in general, decision procedures for \lia have been
proposed that often determine the satisfiability of formulas that
arise from practical verification problems
efficiently~\cite{demoura08}.
\sys assumes access to a decision procedure for \lia, denoted \issat.

An interpolant of mutually inconsistent formulas $\varphi_0$ and
$\varphi_1$ is a \lia formula that explains their inconsistency using
their common vocabulary.
\begin{defn}
  \label{defn:itps}
  For spaces of logical variables $X$ and $Y$,
  $\varphi_0 \in \formulas{X}$ and
  $\varphi_1 \in \formulas{Y}$,
  if $I \in \formulas{X \intersection Y}$ is such that 
  \textbf{(1)} $\varphi_0 \entails I$ and %
  \textbf{(2)} $I, \varphi_1 \entails \false$, then $I$ is an
  \emph{interpolant} of $\varphi_0$ and $\varphi_1$.
\end{defn}
Previous work has introduced interpolating theorem provers that
synthesize interpolants of pairs of mutually-unsatisfiable formulas in
extensions theories used to model program semantics and
specifications~\cite{mcmillan04}.
To present \sys, we assume access to a procedure \solveitp that, given
mutually unsatisfiable \lia formulas $\varphi_0, \varphi_1$, returns
an interpolant of $\varphi_0$ and $\varphi_1$.

\subsubsection{Symbolic representation of program semantics}
\label{sec:symbolic-semantics}
The semantics of $\lang$ can be represented symbolically using \lia
formulas.
In particular, each program store $\sigma \in \stores$ corresponds to
a \lia model over the vocabulary $\vars$, denoted $\modelof{\sigma}$.
For each space of variables $X$, space of indices $I$ and index $i \in
I$, the space of variables $X_i$ denotes a distinct copy of the
variables in $X$.
$X'$ denotes primed copies of $X$, which will typically be used to
model the post-state resulting from an instruction.

For each instruction $\cc{i} \in \instrs$, there is a formula
$\symrelof{\cc{i}} \in \formulas{\vars, \vars'}$ such that for all
stores $\sigma, \sigma' \in \stores$, $(\sigma, \sigma') \in
\transrelof{\cc{i}}$ if and only if $\modelof{\sigma},
\modelof{\sigma'} \sats \symrelof{\cc{i}}$.
A symbolic relation is a formula whose models define pairs of states
from distinct programs.
The space of symbolic relations is denoted $\symrels =
\formulas{\vars_0, \vars_1}$.


\section{Technical Approach}
\label{sec:approach}
In this section, we describe our approach in technical detail.
In \autoref{sec:proof-structure}, we define a class of proof
structures that each represent a product program paired with its
inductive invariants.
In \autoref{sec:solve}, we describe \sys, which given two programs,
attempts to prove or falsify their equivalence by synthesizing such a
proof structure.
In \autoref{sec:discussion}, we state and prove the correctness of
\sys, and compare it to related approaches for proving program
equivalence.
Proofs for each lemma and theorem stated in this section are given in
\autoref{app:proofs}.

\subsection{Proof structures}
\label{sec:proof-structure}
For fixed $\cc{P}_0, \cc{P}_1 \in \lang$, location-pair invariants of
$\cc{P}_0$ and $\cc{P}_1$ describe each pair of runs of $\cc{P}_0$ and
$\cc{P}_1$.
Location-pair invariants are represented as a map from each pair of
control locations to a symbolic relation that describes pairs of
states of $\cc{P}_0$ and $\cc{P}_1$ at the mapped pair of locations.
Let the space of location-pair relations be denoted $\locrels = \locs
\times \locs \to \symrels$.
\begin{defn}
  \label{defn:location-pair-invs}
  Let $I_0, I_1 \in \locrels$ be such that 
  \textbf{(1)} $\params_0 = \params_1 \entails I_0(\initloc, \initloc)
  \lor I_1(\initloc, \initloc)$,
  \textbf{(2)} for each $\cc{i} \in \cc{P}_0$ and $\cc{L} \in
  \locs$, %
  \[ I_0(\prelocof{ \cc{i} }, \cc{L}), %
  \subs{ \symrelof{ \cc{i} } }{ \vars_0, \vars_0' } \entails %
  \replace{ (I_0(\brtargetof{ \cc{i} }, \cc{L} ) \lor 
    I_1(\brtargetof{ \cc{i} }, \cc{L} ) ) }{ \vars_0 }{ \vars_0' } \]
  \textbf{(3)} for each $\cc{L} \in \locs$ and $\cc{i} \in \cc{P}_1$,
  \[ I_1(\cc{L}, \prelocof{ \cc{i} }), %
    \subs{ \symrelof{ \cc{i} } }{ \vars_1, \vars_1' } \entails %
    \replace{ ( I_0(\cc{L}, \brtargetof{ \cc{i} } ) \lor 
      I_1( \cc{L}, \brtargetof{ \cc{i} } ) ) }{ \vars_1 }{ \vars_1' } \]
  \textbf{(4)} $I_0(\finalloc, \finalloc) \entails \retvar_0 =
  \retvar_1$ and $I_1(\finalloc, \finalloc) \entails \retvar_0 =
  \retvar_1$.

  Then $(I_0, I_1)$ are \emph{location-pair invariants} of $\cc{P}_0$
  and $\cc{P}_1$.
\end{defn}
The space of location-pair invariants for $\cc{P}_0$ and $\cc{P}_1$ is
denoted $\locinvs{ \cc{P}_0 }{ \cc{P}_1 }$.

Location-pair invariants for $\cc{P}_0$ and $\cc{P}_1$ define both a
product program $\cc{P}'$ for $\cc{P}_0$ and $\cc{P}_1$, along with
inductive invariants of $\cc{P}'$ that imply that $\cc{P}_0 \equiv
\cc{P}_1$, as described in \autoref{sec:ex-rel-invs}.
Let $(I_0, I_1)$ be location-pair invariants for $\cc{P}_0$ and
$\cc{P}_1$;
the product program $\cc{P}'$ defined by $(I_0, I_1)$ is as follows.
For all $\cc{L}_0, \cc{L}_1 \in \locs$, if $\cc{P}'$ is in a state
that satisfies $I_0(\cc{L}_0, \cc{L}_1)$, then $\cc{P}'$ may choose
$\cc{P}_0$ as its stepping program;
if $\cc{P}'$ is in a state that satisfies $I_1(\cc{L}_0, \cc{L}_1)$,
then $\cc{P}'$ may choose $\cc{P}_1$ as its stepping program.
Otherwise, the next step $\cc{P}'$ is undefined in its current state.

The inductive invariants of $\cc{P}'$ are, for all $\cc{L}_0, \cc{L}_1
\in \locs$, $I_0(\cc{L}_0, \cc{L}_1) \lor I_1(\cc{L}_0, \cc{L}_1)$.
\begin{ex}
  \cc{climbStairs0} and \cc{climbStairs1} have location-pair
  invariants $I_0, I_1 \in \locrels$ that correspond to the product
  program $\cc{climbStairs}'$ and its inductive invariants given in
  \autoref{sec:ex-rel-invs}.
  Key entries in $I_0$ include
  \begin{align*}
    I_0( \cc{2}, \cc{2} ) \equiv\ & \cc{n}_0 = \cc{n}_1 &
    I_0( \cc{5}, \cc{5} ) \equiv\ & \true \\
    I_0( \cc{9}, \cc{9} ) \equiv\ & %
    \cc{i}_0 = \cc{i}_1 + 1 \land \cc{sum}_0 = \cc{count2}_1 \land %
    \cc{n}_0 = \cc{n}_1 &
    I_0(\cc{15}, \cc{15} ) \equiv\ & \cc{result}_0 = \cc{result}_1
  \end{align*}
  $I_1$ at each of the location pairs given above is $\false$.
  At pairs of locations that are not the same line numbers in
  \cc{climbStairs0} and \cc{climbStairs1}, $I_0$ is $\false$ and $I_1$
  is a suitable symbolic relation.
\end{ex}

Location-pair invariants for $\cc{P}_0$ and $\cc{P}_1$ are evidence of
the partial equivalence of $\cc{P}_0$ and $\cc{P}_1$.
\begin{lemma}
  \label{lemma:valid-proof}
  If there are $I_0, I_1 \in \locinvs{ \cc{P}_0 }{ \cc{P}_1 }$, then
  $\cc{P}_0 \equiv \cc{P}_1$.
\end{lemma}

\sys attempts to synthesize location-pair invariants from maps from
pairs of paths to symbolic relations.
Let a \emph{path-pair relation} be a partial map from pairs of paths
to symbolic relations;
i.e., the space of path-pair relations is $\pathpairrels{ \cc{P}_0 }{
  \cc{P}_1 } = \paths{ \cc{P}_0 } \times \paths{ \cc{P}_1 } \partto
\symrels$.
Path-pair relations that \textbf{(1)} are supported by the assumption
that runs of $\cc{P}_0$ and $\cc{P}_1$ begin with equal arguments, %
\textbf{(2)} soundly model steps of execution of $\cc{P}_0$, %
\textbf{(3)} soundly model steps of execution of $\cc{P}_1$, and %
\textbf{(4)} support the conclusion that all modeled pairs of complete
paths end in states with equal return values are path-pair invariants.
\begin{defn}
  \label{defn:path-pair-invs}
  Let $I \in \pathpairrels{ \cc{P}_0 }{ \cc{P}_1 }$ be such that %
  \textbf{(1)} $\params_0 = \params_1 \entails I([ \initloc ], [
  \initloc ] )$;
  \textbf{(2)} for each $p_0 \in \paths{ \cc{P}_0 }$, %
  $\cc{i} \in \cc{P}_0$ and %
  $p_1 \in \paths{ \cc{P}_1 }$ such that $(p_0 \concat \prelocof{
    \cc{i} } \concat \brtargetof{ \cc{i} }, p_1) \in \domain(I)$
  (where for function $f$, $\domain(f)$ denotes the domain of $f$),
  \[ I(p_0 \concat \prelocof{ \cc{i} }, p_1), %
  \subs{ \symrelof{ \cc{i} } }{ \vars_0, \vars_0' } \entails %
  \subs{ I(p_0 \concat \prelocof{ \cc{i} } \concat %
    \brtargetof{ \cc{i} } , p_1) }{ \vars_0', \vars_1 } \]
  \textbf{(3)} for each $p_0 \in \paths{ \cc{P}_0 }$, %
  $p_1 \in \paths{ \cc{P}_1 }$, and %
  $\cc{i} \in \cc{P}_1$ such that $(p_0, p_1 \concat \prelocof{ \cc{i}
  } \concat \brtargetof{ \cc{i} }) \in \domain(I)$,
  \[ I(p_0, p_1 \concat \prelocof{ \cc{i} }), %
  \subs{ \symrelof{ \cc{i} } }{ \vars_1, \vars_1' } \entails %
  \subs{ I(p_0, p_1 \concat \prelocof{ \cc{i} } \concat %
    \brtargetof{ \cc{i} } ) }{ \vars_0, \vars_1' } \]
  \textbf{(4)} for all complete paths $p_0 \in \paths{ \cc{P}_0 }$ and
  $p_1 \in \paths{ \cc{P}_1 }$, $I(p_0, p_1) \entails \retvar_0 =
  \retvar_1$.

  Then $I$ are \emph{path-pair invariants} of $\cc{P}_0$ and
  $\cc{P}_1$.
\end{defn}
The space of path-pair invariants for $\cc{P}_0$ and $\cc{P}_1$ is
denoted $\pathpairinvs{ \cc{P}_0 }{ \cc{P}_1 }$.
For $p_0 \in \paths{ \cc{P}_0 }$ and $p_1 \in \paths{ \cc{P}_1 }$, the
space of path-pair invariants in which $(p_0, p_1)$ is defined is
denoted $\pathpairinvs{ p_0 }{ p_1 }$.

If path-pair invariants $I$ define a product program and inductive
invariants that prove $\cc{P}_0 \equiv \cc{P}_1$, then $I$ are
inductive for $\cc{P}_0$ and $\cc{P}_1$.
For $R \in \pathpairrels{ \cc{P}_0 }{ \cc{P}_1 }$, let $\locrelsof{R}
\in \locrels$ be such that for all $\cc{L}_0 ,\cc{L}_1 \in \locs$,
\[ \locrelsof{R}(\cc{L}_0, \cc{L}_1) = %
\biglor \setformer{ R(p_0 \concat \cc{L}_0, p_1 \concat \cc{L}_1) }{ %
  p_0 \in \paths{ \cc{P}_0 }, %
  p_1 \in \paths{ \cc{P}_1 }, %
  \cc{L}_0, \cc{L}_1 \in \locs, %
  (p_0 \concat \cc{L}_0, p_1 \concat \cc{L}_1) \in \domain(R) } \]
\begin{defn}
  \label{defn:inductive-path-pair-invers}
  For $I \in \pathpairinvs{ \cc{P}_0 }{ \cc{P}_1 }$, %
  if there are $R_0, R_1 \in \pathpairrels{ \cc{P}_0 }{ \cc{P}_1 }$
  such that $I = R_0 \union R_1$ and $(\locrelsof{R_0},
  \locrelsof{R_1})$ are location-pair invariants of $\cc{P}_0$ and
  $\cc{P}_1$ (\autoref{defn:location-pair-invs}), then $I$ are
  \emph{inductive path-pair invariants} for $\cc{P}_0$ and $\cc{P}_1$.
\end{defn}
Inductive path-pair invariants for $\cc{P}_0$ and $\cc{P}_1$ are
evidence of partial equivalence, by \autoref{lemma:valid-proof}.
\sys, given $\cc{P}_0$ and $\cc{P}_1$, attempts to prove $\cc{P}_0
\equiv \cc{P}_1$ by synthesizing inductive path-pair invariants of
$\cc{P}_0$ and $\cc{P}_1$.
\begin{ex}
  The path-pair invariants $I_p$ relating path $p_0$ of
  \cc{climbStairs0} and path $p_1$ of \cc{climbStairs1} (given in
  \autoref{sec:ex-verify}, \autoref{fig:path-pair-invs}) prove their
  partial equivalence.
  $I_p$ cannot be expressed as the union of any two path-pair
  relations $R_0$ and $R_1$ such that $(\locrelsof{R_0},
  \locrelsof{R_1})$ are location-pair invariants, as discussed in
  \autoref{sec:ex-verify}.
  Thus, $I_p$ are not inductive path-pair invariants of
  \cc{climbStairs0} and \cc{climbStairs1}.
\end{ex}

\subsection{Verification algorithm}
\label{sec:solve}
\begin{figure}
  \centering
\begin{minipage}{.49\textwidth} %
  \vspace{0pt}
  \begin{algorithm}[H]
    \SetKwInOut{Input}{Input}
    \SetKwInOut{Output}{Output}
    \SetKwProg{myproc}{Procedure}{}{}
    \Input{$\cc{P}_0, \cc{P}_1 \in \lang$}
    \Output{A decision as to whether $\cc{P}_0 \equiv \cc{P}_1$}
    \myproc{$\sys(\cc{P}_0, \cc{P}_1)$ \label{line:solve-begin}} %
    { \myproc{$\solveaux(I)$ \label{line:aux-begin} }{ %
        \Switch{$\chkinductive(\cc{P}_0, \cc{P}_1, I)$ %
          \label{line:chkind} }{ %
          \lCase{$\isind$}{ %
            \Return{$\true$} \label{line:ret-equiv} %
          } %
          \Case{$p_0 \in \paths{\cc{P}_0}, p_1 \in \paths{\cc{P}_1}$ %
            \label{line:case-cex}}{ %
            \Switch{$\refine(\cc{P}_0, \cc{P}_1, p_0, p_1)$ %
              \label{line:refine} }{ %
              \lCase{$\nonequiv$}{ %
                \Return{$\false$} \label{line:ret-inequiv} %
              } %
              \Case{$I' \in \pathpairinvs{ p_0 }{ p_1 }$}{ %
                \Return{$\solveaux(\mergeinvs(I, I'))$} %
                \label{line:recurse}
              } %
            } %
          } %
        } %
      } \label{line:aux-end} %
      \Return{$\solveaux(\emptyset)$} \label{line:base-call}
    } %
    \caption{%
      \sys: given $\cc{P}_0$ and $\cc{P}_1$, determines if $\cc{P}_0
      \equiv \cc{P}_1$, using procedures \chkinductive and $\refine$,
      which are discussed in \autoref{sec:solve}. }
    \label{alg:sys}
  \end{algorithm}
\end{minipage}
\qquad
\begin{minipage}{0.46\linewidth}
  \begin{algorithm}[H]
    \SetKwInOut{Input}{Input}
    \SetKwInOut{Output}{Output}
    \SetKwProg{myproc}{Procedure}{}{}
    \Input{$\cc{P}_0, \cc{P}_1$ and %
      $I \in \pathpairinvs{ \cc{P}_0 }{ \cc{P}_1 }$. }
    \Output{$\isind$ to denote that some restriction of $I$ are
      inductive path-pair invariants or %
      a pair of paths not defined in $I$.}
    \myproc{$\chkinductive(\cc{P}_0, \cc{P}_1, I)$ %
      \label{line:unsub-begin}} %
    { 
      \myproc{$\chkinductiveaux(\obligations, \discharged)$
        \label{line:unsub-aux-begin}}
      { 
        \lIf{$\obligations = \emptyset$ }{ %
          \Return{$\isind$} \label{line:ret-is-empty}%
        } %
        $((p_0, p_1), \obligations') \assign %
        \remelt(\obligations)$ \label{line:choose-obl} \;
        \If{$(p_0, p_1) \notin \domain(I)$ \label{line:test-def} } %
        { \Return{$(\mincomplete(\cc{P}_0, p_0), %
            \mincomplete(\cc{P}_1, p_1))$} \label{line:ret-cex} }
        $\discharged' \assign \add{\discharged}{(p_0, p_1)}$ %
        \label{line:ext-dis} \;
        $r \assign %
        \chkinductiveaux(\obligations', \discharged')$ %
        \label{line:chkind-rec} \;
        $r_0 \assign \chkinductiveaux( %
        \obligations' \union %
        \exts(\cc{P}_0, p_0) \times \elts{p_1}, \discharged')$ %
        \label{line:chkind-rec-0} \;
        $r_1 \assign \chkinductiveaux( %
        \obligations' \union \elts{p_1} \times \exts(\cc{P}_1, p_1), 
        \discharged')$ %
        \label{line:chkind-rec-1} \; %
        \lIf{$\isdis(I, p_0, p_1, \discharged)$}{ %
          \Return{$r$} \label{line:chkind-ret} %
        } %
        \lElseIf{$\tailof{p_0} = \finalloc$}{ %
          \Return{$r_1$} \label{line:chkind-ret-0} %
        } %
        \lElseIf{$\tailof{p_1} = \finalloc$}{ %
          \Return{$r_0$} \label{line:chkind-ret-1} %
        } %
        \lElse{ %
          \Return{$\chooseres(r_0, r_1)$} %
          \label{line:chkind-ret-choose} 
        } %
      } \label{line:unsub-aux-end} %
      \Return{$\chkinductiveaux( %
        \elts{([ \initloc ], [ \initloc ])}, \emptyset)$} %
      \label{line:unsub-aux-init} \;
    } %
    \caption{$\chkinductive$: given $\cc{P}_0, \cc{P}_1 \in \lang$ and
      $I \in \pathpairinvs{ \cc{P}_0 }{ \cc{P}_1 }$, returns $\isind$
      to denote that some restriction of $I$ are inductive or a pair
      of a paths not defined $I$. }
    \label{alg:chk-ind}
  \end{algorithm}
\end{minipage}
\end{figure}
Pseudocode for the core algorithm implemented by \sys is given in
\autoref{alg:sys}.
The core algorithm is structured as a counterexample-guided refinement
loop analogous to conventional automatic verifiers of safety
properties~\cite{bjorner13,mcmillan06}.
\sys takes $\cc{P}_0, \cc{P}_1 \in \lang$ as input
(\autoref{line:solve-begin}).
\sys defines a procedure \solveaux that, given $I \in \pathpairinvs{
  \cc{P}_0 }{ \cc{P}_1 }$, attempts to determine if $\cc{P}_0 \equiv
\cc{P}_1$ by constructing inductive path-pair invariants from $I$
(\autoref{line:aux-begin}---\autoref{line:aux-end}).
\sys runs \solveaux on the empty path-pair relation and returns the
result (\autoref{line:base-call}).

\solveaux, given path-pair invariants $I$ (\autoref{line:aux-begin}),
first runs a procedure $\chkinductive$ on $\cc{P}_0$, $\cc{P}_1$, and
$I$ (\autoref{line:chkind}).
If $\chkinductive$ returns value $\isind$ to denote that some
restriction of $I$ are inductive path-pair invariants of $\cc{P}_0$
and $\cc{P}_1$, then $\solveaux$ returns $\true$, to denote $\cc{P}_0
\equiv \cc{P}_1$ (\autoref{line:ret-equiv}).
Otherwise, if $\chkinductive$ returns a pair of paths $p_0$ and $p_1$
that are not defined in $I$ (\autoref{line:case-cex}), then
$\solveaux$ runs a procedure $\refine$ on $\cc{P}_0$, $\cc{P}_1$,
$p_0$ and $p_1$ (\autoref{line:refine}).
If $\refine$ returns that $p_0 \not\equiv p_1$, then $\solveaux$
returns $\false$, to denote $\cc{P}_0 \not\equiv \cc{P}_1$
(\autoref{line:ret-inequiv}).

Otherwise, if $\refine$ returns $I' \in \pathpairinvs{p_0}{p_1}$, then
$\solveaux$ runs $\mergeinvs$ on $I$ and $I'$ to obtain path-pair
invariants defined over all pairs of paths defined in $I$ or $I'$,
recurses on the result, and returns the result of the recursion
(\autoref{line:recurse}).
$\mergeinvs$ returns $I'' \in \pathpairinvs{ \cc{P}_0 }{ \cc{P}_1 }$
such that for each $p_0 \in \paths{ \cc{P}_0 }$ and $p_1 \in \paths{
  \cc{P}_1 }$,
if $(p_0, p_1) \in \domain(I) \setminus \domain(I')$, then $I''(p_0,
p_1) = I(p_0, p_1)$;
if $(p_0, p_1) \in \domain(I') \setminus \domain(I)$, then $I''(p_0,
p_1) = I'(p_0, p_1)$;
otherwise, $I''(p_0, p_1) = I(p_0, p_1) \land I'(p_0, p_1)$.

\subsubsection{Finding path-pair invariants using \refine}
\label{sec:find-path-invs}
$\refine$, given $\cc{P}_0, \cc{P}_1 \in \lang$, $p_0 \in \paths{
  \cc{P}_0 }$, and $p_1 \in \paths{ \cc{P}_1 }$, either returns
path-pair invariants of $p_0$ and $p_1$ or determines that $p_0
\not\equiv p_1$.
$\refine$ attempts to find invariants of each $p_0' \in \prefixesof{
  p_0 }$ paired with each $p_1' \in \prefixesof{ p_1 }$ as the
interpolant of %
\textbf{(1)} the disjunction of path-pair invariants describing all
pairs of states immediately before $\cc{P}_0$ and $\cc{P}_1$ take a
final step to complete $p_0'$ and $p_1'$ and %
\textbf{(2)} a formula describing all pairs of states at $p_0'$ and
$p_1'$ from which the remainder of $p_0$ and $p_1$ result in states
with non-equal return values.

$\refine$ performs the following procedure.
For each $p_0' \in \prefixesof{ p_0 }$, let there be a distinct copy
of $\vars$ denoted $\vars[ p_0' ]$.
Let $\remainctr{ \cc{P}_0 }{ p_0' } \in \formulas{\bigunion_{p \in
    \prefixesof{p_0}} \vars[p]}$ be the conjunction of semantic
constraints from all steps following in $p_0$ following $p_0'$:
\[ \bigland_{
  \substack{
    p_0'' \in \prefixesof{ p_0 }, \\
    \cc{L}, \cc{L}' \in \locs, \\
    p_0'' \concat \cc{L} \concat \cc{L}' \in %
    \subrange(p_0', p_0) } } %
\symrelof{ \instrat{ \cc{P}_0 }(\cc{L}, \cc{L}')}( %
\vars[ p_0'' \concat \cc{L} ], %
\vars[ p_0'' \concat \cc{L} \concat \cc{L} '])
\]
For each $p_1' \in \prefixesof{ p_1 }$, $\remainctr{ \cc{P}_1 }{ p_1'
}$ is defined similarly.

$\refine$ first determines if $p_0$ and $p_1$ are equivalent by
running $\issat$ on a formula $\noneqpaths{ p_0 }{ p_1 }$ for which
each model corresponds to a run of $p_0$ paired with a run of $p_1$
that start with equal parameter values and complete with unequal
return values.
I.e., $\noneqpaths{ p_0 }{ p_1 }$ is:
\[
\params_0[ [ \initloc ] ] = \params_1[ [ \initloc ] ] \land 
\remainctr{\cc{P}_0}{[ \initloc ]} \land \remainctr{\cc{P}_1}{[
  \initloc ]} \land
\retvar_0[ p_0 ] \not= \retvar_1[ p_1 ] \]
If $\noneqpaths{p_0}{p_1}$ is satisfiable, then $\refine$ returns
$\nonequiv$.
\begin{ex}
  To determine if $p_0$ from \cc{climbStairs0} and $p_1$ from
  \cc{climbStairs1} (see \autoref{sec:ex-verify}) are partially
  equivalent, \sys determines the satisfiability of the following
  formula:
  \begin{align*}
    \cc{n}_0 = \cc{n}_1 \land %
    \remainctr{\cc{climbStairs0}}{[ \cc{2} ]} \land %
    \remainctr{\cc{climbStairs1}}{[ \cc{2} ]} \land %
    \cc{result}_0 \not= \cc{result}_1
  \end{align*}
  \sys uses \issat to determine that the above formula is
  unsatisfiable, and thus that the $p_0 \equiv p_1$.
\end{ex}

If $\noneqpaths{p_0}{p_1}$ is unsatisfiable, then $p_0 \equiv p_1$.
In such a case, $\refine$ computes, for each $p_0' \in
\prefixesof{p_0}$ paired with each $p_1' \in \prefixesof{p_1}$, a
path-pair invariant $I(p_0', p_1')$ as an interpolant of two formulas.
The first formula, referred to as the $\mathsf{PreCtr}[ p_0', p_1' ]$,
is determined by the form of $p_0'$ and $p_1'$.
$\mathsf{PreCtr}[ [ \initloc ], [ \initloc] ]$ is
\[ \params_0[ [ \initloc ] ] = \params_1[ [ \initloc] ]
\]
For $p_0' \in \prefixesof{p_0}$ and $\cc{L} \in \locs$ such that $p_0'
\concat \cc{L} \in \prefixesof{p_0}$, $\mathsf{PreCtr}[p_0' \concat
\cc{L}, [ \initloc ] ]$ is
\[  I(p_0', [ \initloc]) \land %
\symrelof{\instrat{\cc{P}_0}(\vars_0[ p_0' ], \vars_0[p_0' \concat \cc{L} ])}
\]
For $p_1' \in \prefixesof{p_1}$ and $\cc{L} \in \locs$ such that $p_1'
\concat \cc{L} \in \prefixesof{p_1}$, $\mathsf{PreCtr}[ [ \initloc ],
p_1' \concat \cc{L} ]$ is
\[ I([ \initloc ], p_1') \land %
\symrelof{\instrat{\cc{P}_1}(\vars_1[ p_1' ], \vars_1[p_1' \concat \cc{L} ])}
\]
For $p_0' \in \prefixesof{p_0}$, %
$\cc{L}_0 \in \locs$ such that $p_0' \concat \cc{L}_0 \in
\prefixesof{p_0}$, $p_1' \in \prefixesof{p_1}$, and %
$\cc{L}_1 \in \locs$ such that $p_1' \concat \cc{L}_1 \in
\prefixesof{p_1}$, $\mathsf{PreCtr}[ p_0' \concat \cc{L}_0, p_1'
\concat \cc{L}_1 ]$ is
\[ 
  ( I(p_0', p_1') \land %
  \subs{ \symrelof{ \instrat{ \cc{P}_0 } } }{ %
      \vars_0[ p_0' ], \vars_0[ p_0' \concat \cc{L}_0 ] } ) \lor 
  (I(p_0', p_1') \land %
  \subs{\symrelof{ \instrat{ \cc{P}_1 } } }{
    \vars_1 [ p_1' ], \vars_1[ p_1' \concat \cc{L}_1 ] } ) 
\]

The second formula, referred to as the post-constraint
$\postctr{p_0' }{ p_1' }$ is
\[ \remainctr{ \cc{P}_0 }{ p_0' } \land %
\remainctr{ \cc{P}_1 }{ p_1' } \land %
\retvar_0[ p_0 ] \not= \retvar_1[ p_1 ]
\]
\begin{ex}
  \sys, given path $p_0$ of \cc{climbStairs0} and path $p_1$ of
  \cc{climbStairs} (see \autoref{sec:ex-verify}), synthesizes the
  path-pair invariants of each prefix of $p_0$ paired with each prefix
  of $p_1$ using a series of queries to an interpolating theorem
  prover.
  E.g., in order to synthesize the pair-pair invariant that relates
  prefix $[ \cc{2}, \cc{9} ]$ of $p_0$ to prefix $[ \cc{2}, \cc{9},
  \cc{9} ]$ of $p_1$, \sys synthesizes a pre-constraint consisting of
  the disjunction of \textbf{(1)} the path-pair invariant for $([
  \cc{2} ], [ \cc{2}, \cc{9} ])$ combined with the semantics of
  \cc{climbStairs0} stepping from \cc{2} to \cc{9} and %
  \textbf{(2)} the path-pair invariant for $([ \cc{2}, \cc{9} ], [
  \cc{2}, \cc{9} ])$ combined with the semantics of \cc{climbStairs1}
  taking a step from \cc{9} to \cc{9}.
  \sys computes the invariants for both of the pairs of paths given
  above from previous interpolation queries.

  \sys constructs a post-constraint consisting of the conjunction of
  \textbf{(1)} $\remainctr{ \cc{P}_0 }{ [ \cc{2}, \cc{9} ] }$, which
  models \cc{climbStairs0} stepping from \cc{9} to \cc{9} and then
  from \cc{9} to \cc{15}, and %
  \textbf{(2)} $\remainctr{ \cc{P}_1 }{ [ \cc{2}, \cc{9}, \cc{9} ]}$,
  which models \cc{climbStairs1} stepping from \cc{9} to \cc{9} and
  then from \cc{9} to \cc{15}, and %
  \textbf{(3)} $\cc{result}_0 \not= \cc{result}_1$.

  One interpolant of the pre-constraint and post-constraint given
  above is $I(\cc{9}, \cc{9})$, the invariant for location \cc{9} in
  \cc{climbStairs0} and \cc{9} in \cc{climbStairs1} that is also a
  path-pair invariant for paths $[ \cc{2}, \cc{9} ]$ and $[ \cc{2},
  \cc{9}, \cc{9} ]$, as depicted in \autoref{fig:path-pair-invs}.
\end{ex}

For each $p_0' \in \prefixesof{p_0}$, $p_1' \in \prefixesof{p_1}$,
$I(p_0', p_1')$ is the interpolant of $\mathsf{PreCtr}[ p_0', p_1']$
and $\postctr{p_0'}{p_1'}$.
The entries of $I$ can be computed in any ordering of the pairs of
prefixes of $p_0$ and $p_1$ that respects the prefix ordering of both
$p_0$ and $p_1$.
$\refine$ returns the path-pair relations $I' \in \pathpairinvs{ p_0
}{ p_1 }$ such that for each $p_0' \in \prefixesof{ p_0 }$ and $p_1'
\in \prefixesof{ p_1 }$, $I'(p_0', p_1') = \subs{ I(p_0', p_1') }{
  \vars_0, \vars_1 }$.

The correctness of \sys is partially established by the fact that
\refine returns path-pair relations exactly when it is given a pair of
paths that are equivalent.
\begin{lemma}
  \label{lemma:find-inv-corr}
  For all $p_0 \in \paths{ \cc{P}_0 }$ and $p_1 \in \paths{ \cc{P}_1
  }$, if $p_0 \equiv p_1$, then $\refine(\cc{P}_0, \cc{P}_1, p_0,
  p_1) \in \pathpairinvs{p_0}{p_1}$.
  Otherwise, $\pathpairinvs{p_0}{p_1} = \nonequiv$.
\end{lemma}

\subsubsection{Finding inductive path-pair invariants using \chkinductive}
\label{sec:chk-inductive}
\autoref{alg:chk-ind} contains pseudocode for \chkinductive.
\chkinductive, given $\cc{P}_0$, $\cc{P}_1 \in \lang$ and path-pair
invariants $I \in \pathpairrels{ \cc{P}_0 }{ \cc{P}_1 }$
(\autoref{line:unsub-begin}), returns either %
\textbf{(1)} the value $\isind$ to denote that some restriction of $I$
is inductive path-pair invariants of $\cc{P}_0$ and $\cc{P}_1$, or %
\textbf{(2)} a pair of paths of $\cc{P}_0$ and $\cc{P}_1$ that have no
invariant in $I$.
\chkinductive defines a procedure \chkinductiveaux
(\autoref{line:unsub-aux-begin}---\autoref{line:unsub-aux-end}) that
takes two sets of pairs of paths: %
\textbf{(1)} \emph{obligation} pairs $\obligations$ and %
\textbf{(2)} \emph{discharged} pairs $\discharged$.
\chkinductiveaux returns either %
\textbf{(1)} the value $\isind$ to denote that $\cc{P}_0$ and
$\cc{P}_1$ have inductive path-pair invariants defined by $I$
restricted to some set of path-pairs that contains $\discharged \union
\obligations$ or %
\textbf{(2)} a pair of paths that are an extension of some pair in
$\obligations$ that have no invariant in $I$.
\chkinductive runs \chkinductiveaux on an initial set of obligations
that contains only $([ \initloc ], [ \initloc ])$ and an empty set of
discharged path pairs, and returns the result
(\autoref{line:unsub-aux-init}).

\chkinductiveaux first tests if $\obligations$ is empty, and if so
returns $\isind$ (\autoref{line:ret-is-empty}).
Otherwise, if $\obligations$ is not empty, then \chkinductiveaux
chooses and removes a path-pair $(p_0, p_1)$ from $\obligations$
(\autoref{line:choose-obl}).
\chkinductiveaux then tests if $(p_0, p_1)$ is undefined in $I$
(\autoref{line:test-def}) and, if so, returns a pair of a
minimum-length complete extensions of $p_0$ and $p_1$
(\autoref{line:ret-cex}).

Otherwise, \chkinductiveaux extends $\discharged$ to contain $(p_0,
p_1)$ to form $\discharged'$ (\autoref{line:ext-dis}), and computes
the result of recursing on $\discharged'$ on three distinct sets of
obligations:
\textbf{(1)} $\obligations'$, the result of which is stored in $r$
(\autoref{line:chkind-rec});
\textbf{(2)} $\obligations'$ extended with all control successors in
$\cc{P}_0$ of $(p_0, p_1)$ (denoted $\exts(\cc{P}_0, p_0)$), the
result of which is stored in $r_0$ (\autoref{line:chkind-rec-0});
\textbf{(3)} $\obligations'$ extended with all control successors in
$\cc{P}_1$ of $(p_0, p_1)$ (denoted $\exts(\cc{P}_1, p_1)$), the
result of which is stored in $r_1$ (\autoref{line:chkind-rec-1}).

\chkinductiveaux tests if $I(p_0, p_1)$ entails the invariant in $I$
for some discharged pair of paths with the same final locations by
computing:
\[ \isdis(I, p_0, p_1, \discharged) = %
\biglor \setformer{  %
  I(p_0', p_1') \entails I(p_0, p_1) }{ %
  (p_0', p_1') \in \discharged, %
  \tailof{ p_0 } = \tailof{ p_0' }, \tailof{ p_1 } = \tailof{ p_1' } } \]
If $\isdis(I, p_0, p_1, \discharged)$ holds, then \chkinductiveaux
returns $r$ (\autoref{line:chkind-ret}).
Otherwise, if only $p_0$ is a complete path, then \chkinductiveaux
returns $r_1$ (\autoref{line:chkind-ret-0}).
Otherwise, if only $p_1$ is a complete path, then \chkinductiveaux
returns $r_0$ (\autoref{line:chkind-ret-1}).
Otherwise, \chkinductiveaux runs a procedure $\chooseres$ on $r_0$ and
$r_1$ (\autoref{line:chkind-ret-choose}).
If either $r_0 = \isdis$ or $r_1 = \isdis$, then $\chooseres$ returns
$\isdis$; 
otherwise, $\chooseres$ returns either result as a complete pair of
paths undefined in $I$ (\autoref{line:chkind-ret-choose}).
\begin{ex}
  \label{ex:cons-path-pair-invs}
  The path-pair invariants $I_p$ described in \autoref{sec:ex-verify}
  are path-pair invariants of $p_0$ of \cc{climbPaths0} and $p_1$ of
  \cc{climbPaths1}.
  However, no restriction of $I_p$ are inductive path-pair invariants
  of \cc{climbStairs0} and \cc{climbStairs1}.
  When \sys inspects $I_p$ to determine if some restriction of $I_p$
  are inductive path-pair invariants, it determines that they are not
  inductive.

  In particular, when \sys first considers the pair of paths
  consisting of only the entry locations $[ \cc{2} ]$ and $[ \cc{2}
  ]$, it does not contain any pair of paths in the set $\discharged$.
  Therefore, \sys only determines that $I_p$ have an inductive
  restriction its recursive call succeeds on either all extensions of
  the pair in \cc{climbStairs0} or \cc{climbStairs1}.
  However, the extensions of the pair in \cc{climbStairs0} include the
  pair of paths $([ \cc{2}, \cc{5} ], [ \cc{2} ])$, and the extensions
  of the pair in \cc{climbStairs1} include the pair of paths $([ \cc{
    2} ], [ \cc{2}, \cc{5} ])$.
  $I_p$ does not define path-pair invariants for either pair of paths.

  \sys therefore returns a pair of complete paths $p_0$ and $p_1$ that
  includes \cc{5} in \cc{climbStairs0} or line \cc{5} in
  \cc{climbStairs1}.
  \sys then synthesizes path-path invariants $I_p''$ for $(p_0, p_1)$
  and $(p_0', p_1')$, as described in \autoref{sec:ex-verify}.
  When \sys calls \chkinductive on $I_p''$, \chkinductive determines
  that some restriction of $I_p''$ are inductive, and thus that
  $\cc{climbStairs0} \equiv \cc{climbStairs1}$.
\end{ex}

The correctness of \sys is partially established partially by the fact
that \chkinductive returns $\isind$ only when given path-pair
invariants that for which some restriction is inductive.
\begin{lemma}
  \label{lemma:find-ind-corr}
  For $I \in \pathpairinvs{ \cc{P}_0 }{ \cc{P}_1 }$, if
  $\chkinductive(\cc{P}_0, \cc{P}_1, I) = \isind$, then some
  restriction of $I$ are inductive path pair invariants of $\cc{P}_0$
  and $\cc{P}_1$.
\end{lemma}

\subsection{Discussion}
\label{sec:discussion}
In this section, we discuss several key properties of \sys.
In \autoref{sec:corr}, we establish \sys's correctness.
In \autoref{sec:self-composition}, we compare to \sys a technique for
proving partial equivalence given in previous work, self-composition.
In \autoref{sec:practice}, we describe challenges to designing a
practical implementation of \sys.
\subsubsection{Correctness}
\label{sec:corr}
Whenever \sys returns a definite result, the result is correct.
\begin{thm}
  \label{thm:soundness}
  For all $\cc{P}_0, \cc{P}_1 \in \lang$, %
  if $\sys(\cc{P}_0, \cc{P}_1)$ is defined, then $\cc{P}_0 \equiv
  \cc{P}_1$ if and only if $\sys(\cc{P}_0, \cc{P}_1) = \true$.
\end{thm}
Because determining partial program equivalence is, in general,
undecidable, \sys is not total: i.e., there are pairs of programs on
which \sys will not terminate.

\sys as presented in \autoref{alg:sys}, given $\cc{P}_0, \cc{P}_1 \in
\lang$, returns a Boolean decision as to whether $\cc{P}_0 \equiv
\cc{P}_1$.
\sys can be directly extended so that if it determines that $\cc{P}_0
\equiv \cc{P}_1$, then it returns inductive path-pair invariants of
$\cc{P}_0$ and $\cc{P}_1$.
In particular, \chkinductive (\autoref{alg:chk-ind}) is extended so
that given path-pair invariants $I$, if it determines that some
restriction of $I$ are inductive path-pair invariants of $\cc{P}_0$
and $\cc{P}_1$, then it returns the restrictions of $I$ that define
location-pair invariants of $\cc{P}_0$ and $\cc{P}_1$.
In such a case, \sys directly returns restrictions obtained from
\chkinductive.

In order to return such restrictions of $I$, \chkinductive maintains,
in addition to the set of obligation path-pair $\obligations$,
\emph{two} sets of discharged pairs of paths, denoted $\discharged_0$
and $\discharged_0$.
When \chkinductive calls itself on pairs of paths constructed from
extensions of $p_0$ in $\cc{P}_0$ (\autoref{line:chkind-rec-0}), it
extends $\discharged_0$ to contain $(p_0, p_1)$.
When \chkinductive calls itself on pairs of paths constructed from
extensions of $p_1$ in $\cc{P}_1$ (\autoref{line:chkind-rec-1}), it
extends $\discharged_1$ to contain $(p_0, p_1)$.
When \chkinductive determines if a given pair of paths $(p_0, p_1)$
has an invariant that is entailed by an invariant that has been
previously discharged by computing the predicate $\isdis$, it
enumerates over $\discharged_0 \union \discharged_1$.

\sys can also be directly extended so that if it determines that
$\cc{P}_0 \not\equiv \cc{P}_1$, then it returns a common input on
which $\cc{P}_0$ and $\cc{P}_1$ generate different final values.
To do so, $\refine$ is extended so that when it is given paths $p_0$
and $p_1$ such that $\noneqpaths{p_0}{p_1}$
(\autoref{sec:find-path-invs}) is satisfiable, $\refine$ returns one
of its models, which is then returned directly by \sys as a pair of
runs from a common input that results in unequal return values.

\subsubsection{Comparison to sequential composition}
\label{sec:self-composition}
Previous work has proposed several approaches for automatically
determining the partial equivalence of programs.
One approach that, given programs $P_0$ and $P_1$, constructs the
\emph{self-composition} of $P_0$ and $P_1$, which is a program that
passes the same inputs to $P_0$ and $P_1$, stores their results, and
asserts that the results are equal~\cite{barthe04,terauchi05}.
Such an approach has potential applications for verifying that a
program satisfies a desired information-flow property, can be
formulated as proving that when a program is given two inputs with
equivalent publicly-visible components, it generates outputs with
equivalent publicy-visible components.
However, such an approach typically cannot be applied to prove that
two programs are partially equivalent, because it requires a safety
prover to infer a summary for each of $P$ and $Q$ that precisely
describes their functionality.
Most model checkers use logics that are combinations of the
quantifier-free fragments of linear arithmetic, uninterpreted
functions, and arrays, which cannot express such summaries.
In particular, neither \cc{climbStairs} solutions given in
\autoref{sec:overview}, nor the solutions that we describe in
\autoref{sec:evaluation} can be precisely summarized in such theories.

\subsubsection{Practical design}
\label{sec:practice}
In \autoref{sec:semantics}, we defined the state space of a \lang
program to be a map from program variables to integer values.
Our prototype implementation of \sys can take as input programs
represented in JVM bytecode, which use instructions that also
dynamically allocate, load from, and store to dynamic memory and
arrays.
In order to support programs that execute such instructions, \sys uses
formulas that axiomatize the semantics of each instruction in the
combination of the theory of linear arithmetic with the theory of
arrays.
Formulas in such theories can also be used to define equivalent
initial or final states that contain linked data structures and
arrays.

The key properties that must be satisfied by a theory $\mathcal{T}$
used by \sys to axiomatize instructions are that %
\textbf{(1)} \sys must have access to an interpolating theorem prover
for $\mathcal{T}$, which it uses to generate path-pair invariants
(\autoref{sec:find-path-invs});
\textbf{(2)} \sys must have access to an automatic decision procedure
for $\mathcal{T}$, used by \sys to check entailments between pair-pair
invariants of different path pairs in \autoref{alg:chk-ind},
\autoref{line:chkind-ret}.

In \autoref{sec:solve}, we described \sys as using several procedures
that were described only at the level of their interface, not their
implementation.
In particular, for fixed $\cc{P} \in \lang$, the procedure
\mincomplete, given $p \in \paths{ \cc{P} }$, returns a complete
extension of $p$.
In general, a control path may have infinitely many complete
extensions.
Our prototype implementation of \sys chooses a complete extension of
minimum length, using breadth-first search.

The procedure \chooseres, used in \chkinductiveaux
(\autoref{sec:chk-inductive}), given two results of recursive calls to
\chkinductiveaux---each of which may be either $\isind$ or a pair of
control paths---returns a final result for the \chkinductiveaux.
Our prototype implementation of \sys, given $\isind$ as either one of
its arguments, always returns $\isind$.
Given two pairs of paths, it always returns the pair with the shortest
combined length.
Other feasible implementations of \sys could be defined by alternative
implementations of \mincomplete and \chooseres that choose paths using
alternative criteria explored by software model checkers for safety
properties.

\chkinductive, given path-pair invariants $I$, can in general execute
in time exponential in the length of the minimal pair of paths not
defined by $I$, as a result of the fact that in each iteration, it
may attempt to find inductive path-pair invariants by extending a path
in $\cc{P}_0$ or $\cc{P}_1$.
Our prototype implementation of \chkinductive lazily call itself
recursively, based on the results of evaluating the predicate $\isdis$
and recursive calls.
The prototype also memoizes sets of obligations and discharged pairs
considered.
While this optimization does not improve \chkinductive's performance
in the worst case, in practice, it causes \chkinductive to perform
significantly more effectively than a conventional inductiveness check
on practical pairs of programs (see \autoref{sec:evaluation}).


\section{Evaluation}
\label{sec:evaluation}
We performed an empirical evaluation of \sys to answer the following
questions:
\textbf{(1)} Can \sys verify the partial equivalence of programs
written independently that implement distinct, subtle algorithms?
\textbf{(2)} Can \sys verify the partial equivalence of programs
written by a wide set of independent programmers?
\textbf{(3)} Can \sys verify equivalence of programs more effectively
than self-composition technical that using generic solver?

To answer the above experimental questions, we implemented \sys as a
partial-equivalence verifier for programs represented in JVM bytecode.
While we presented \sys in \autoref{sec:approach} as a verifier for
programs whose instructions are defined in the theory of linear
arithmetic, the actual implementation models core JVM language
features, including arrays and objects, using the combined theory of
linear arithmetic, arrays, and uninterpreted functions (\auflia).
The only requirement imposed by \sys on the logic for expressing
program semantics is that the logic has \textbf{(1)} an effective
decision procedure, which \sys uses to check entailment over unknown
predicates (\autoref{sec:chk-inductive}), and \textbf{(2)} an
effective procedure that constructs interpolants, which \sys uses to
synthesize path-pair invariants (\autoref{sec:find-path-invs}).
Both operations are supported by the Z3 interpolating theorem
prover~\cite{z3}, which is used in our implementation.
We applied \sys to attempt to prove partial equivalence of 369 pairs
of programs submitted by independent programmers as solutions to
problems hosted on the online coding platforms
Leetcode~\cite{leetcode} and CodeChef~\cite{codechef}.

In short, our experiments answer the above questions positively: \sys
was able to prove the partial equivalence of an overwhelming majority
of pairs of programs to which it was applied.
\sys consistently proved the partial equivalence of programs more
efficiently than self-composition technical that using generic solver.
The results indicate that \sys can synthesize proofs of partial
equivalence effectively enough to be used as an educational aid, or as
an underlying engine for other educational aids, such as
autograders~\cite{singh13}.

\subsection{Experimental procedure}
\label{sec:exp-procedure}
\sys takes as input \textbf{(1)} two programs $P_0$ and $P_1$, each
represented as a JVM bytecode module.
If \sys determines $P_0 \equiv P_1$, then it outputs the relational
invariants of $P_0$ and $P_1$ as the proof.
If \sys determines that $P_0 \not\equiv P_1$, it generates a pair of
runs from $P_0$ and $P_1$ from a common input that result in outputs
that are not equivalent.
\sys is implemented in $4,932$ lines of Java source code.
\sys uses the Soot analysis framework~\cite{soot} to construct the
control-flow graph of given programs, and uses the Z3 interpolating
theorem prover~\cite{z3} to synthesize path-pair invariants (see
\autoref{sec:find-path-invs}).

We collected as benchmarks programs submitted as solutions to problems
posted on the coding platforms LeetCode and CodeChef.
Each problem has over 200 posts in its discussion thread.
To determine if \sys can synthesize proofs of equivalence for many
programs written independently by programmers with a variety of
backgrounds, we collected $369$ pairs of solutions of $14$ different 
programming exercises on LeetCode and CodeChef.
We ran \sys to determine the equivalence of each pairs of solutions.
To show the ability of \sys can synthesize proofs of equivalence
across programs that implement subtle algorithms, we presents four
pairs of solutions submitted for five challenge problems hosted on
LeetCode and CodeChef, in addition to the pair of solutions to
\cc{climbStairs} presented in \autoref{sec:overview}.
The results of running \sys on these benchmarks are described in
detail in \autoref{sec:leetcode}.

The current version of \sys cannot prove equivalence of the vast
majority of solutions on such sites, as proofs of their equivalence
require either quantified invariants over arrays or expressive heap
invariants.
While \sys can model the semantics of such programs accurately,
inferring sufficient invariants over data with such structure is
itself an ongoing topic of research.
We believe that combining \sys with such approaches is an encouraging
direction for future research.

In order to evaluate the ability of \sys to prove equivalence compared
to previous completely-automatic approaches, we implemented an
equivalence verifier, named \baseLine, that uses self-composition
(described in \autoref{sec:self-composition}), to reduce equivalence
verification to safety verification, and apply the best known
techniques for safety verification.
\baseLine, given programs $P_0$ and $P_1$ constructs systems of
\emph{constrained Horn clauses}~\cite{bjorner13} $S_0$ and $S_1$ that
model all executions of $P_0$ and $P_1$.
\baseLine extends $S_0$ and $S_1$ to form a CHC system $S'$ for which
each solution corresponds to invariants of the self-composition of
$P_0$ and $P_1$ that prove their equivalence.
\baseLine then gives $S'$ to \duality, a competitive CHC solver
implemented within the \cc{z3} automated theorem prover.

Verifying equivalence of programs $P_0$ and $P_1$ can be reduced to
verifying safety only if the $P_0$ and $P_1$ read input and write
output to vectors of scalar data, not streams.
As a result, we applied \baseLine to attempt to verify the equivalence
of only programs that operate on scalar data.
Such programs coincided exactly with the programs that we found on
LeetCode.

Both \sys and \baseLine were run on a machine with 16 1.4 GHz
processors and 128 GB of RAM.
The current implementation of \sys uses a single thread.
The implementation is publicly available~\cite{pequod-src}.
All benchmarks are publicly available at references provided in this
paper.
All benchmarks were posted publicly by their programmers, and we have
anonymized the sources of individual programs when referring to them
in our results.
We are working with the administrators of the coding platforms to
potentially redistribute the collected solutions as a standard set of
benchmarks for the verification community.

\subsection{Equivalent solutions of challenge problems}
\label{sec:leetcode}
In this section, we use example solutions from four challenge
problems on LeetCode and CodeChef to illustrate \sys's ability to 
synthesize proofs of equivalence of subtle implementations.
In the relational invariants given for each pair of programs
discussed, variables from the first programs (whose name ends with
\cc{0}) are subscripted $0$ and variables from the second program
(whose name ends with \cc{1}) are subscripted $1$.

\begin{figure}[t]
    \begin{minipage}[b]{0.2\linewidth}
      \input{add-digits-0.java}
      \caption{\cc{addDigits0}: a solution posted for the
        Add Digits problem.
        \label{fig:add-digits-0} }
    \end{minipage}
    \qquad
    \begin{minipage}[b]{0.2\linewidth}
      \input{add-digits-1.java}
      \caption{\cc{addDigits1}: an alternative solution posted for the
        Add Digits problem. \label{fig:add-digits-1} } %
    \end{minipage}
    \qquad
    \begin{minipage}[b]{0.2\linewidth}
      \input{trailing-zeroes-0.java}
      \caption{\cc{trailing0s0}: a solution posted for the
        Trailing Zeroes problem. \label{fig:trailing-zeroes-0} } %
    \end{minipage}
    \qquad
    \begin{minipage}[b]{0.2\linewidth}
      \input{trailing-zeroes-1.java}
      \caption{\cc{trailing0s1}: an alternative solution posted
        for the Trailing Zeroes problem.\label{fig:trailing-zeroes-1}
      } %
    \end{minipage}
\end{figure}

The Add Digits Problem~\cite{AddDigits} is to take a non-negative
integer in variable \cc{num} and return sum of all of the digits in
\cc{num} modulo $9$.
\sys proves that solutions \cc{addDigits1}
(\autoref{fig:add-digits-1}) and \cc{addDigits0}
(\autoref{fig:add-digits-0}) are partially equivalent by synthesizing
the following relational invariant the head of the loop of
\cc{addDigits0} and the end of \cc{addDigits1}:
\[ \cc{result}_0 = \cc{num}_1 - 9\ ((\cc{num}_1 - 1) / 9)
\]

The Trailing Zeroes Problem~\cite{TrailingZeroes} is, given a
non-negative integer $n$, to to return the number of zero digits that
occur before the least-significant non-zero digit in $n!$.
\sys proves that solutions \cc{TrailingZeroes0}
(\autoref{fig:trailing-zeroes-0}) and \cc{TrailingZeroes1}
(\autoref{fig:trailing-zeroes-1}) are equivalent by synthesizing the
following relational invariant over their loop heads:
\[ \cc{sum}_0 = \cc{x}_1 \land \cc{n}_0 = \cc{y}_1 / 5 \land %
(\cc{n}_0 \geq 0 \lor \cc{y}_1 \geq 0) %
\]

\begin{figure}[t]
  \begin{minipage}{0.18\linewidth}
    \input{reverse-integer-0.java}    
    \caption{\cc{reverse0}: a solution provided for the Reverse
      Integer Problem on LeetCode.}
    \label{fig:reverse-0}
  \end{minipage}
  \qquad
  \begin{minipage}{0.18\linewidth}
    \input{reverse-integer-1.java}    
    \caption{\cc{reverse1}: an alternative solution provided for the
      Reverse Integer Problem on LeetCode.}
    \label{fig:reverse-1}
  \end{minipage}
  \qquad
  \begin{minipage}{0.22\linewidth}
    \input{FLOW001-0.java}    
    \caption{\cc{FLOW001_0}: a solution provided for the \cc{FLOW001}
      problem on CodeChef.}
    \label{fig:flow-0}
  \end{minipage}
  \qquad
  \begin{minipage}{0.22\linewidth}
    \input{FLOW001-1.java}    
    \caption{\cc{FLOW001_0}: an alternative solution provided for the
      \cc{FLOW001} problem on CodeChef.}
    \label{fig:flow-1}
  \end{minipage}
\end{figure}

The Reverse Integer Problem~\cite{ReverseInteger} is to take a
non-negative integer $n$ and return an integer that consists of the
digits in $n$ in reversed order.
\sys proves that solutions \cc{reverse0} (\autoref{fig:reverse-0}) and
\cc{reverse1} (\autoref{fig:reverse-1}) are partially equivalent by
synthesizing the following relational invariant over their loop heads:
\[ \cc{x}_0 = \cc{x}_1 \land %
(\cc{x}_0 \geq 0 \lor \cc{x}_1 \geq 0) \land %
\cc{res}_0 = \cc{rev}_1
\]

The Flow-001 Problem~\cite{FLOW001} is to read a non-negative integer
$T$, then read $T$ pairs of integers, printing the sum of each pair of
integers.
\sys proves that two solutions given for the Flow-001 problem,
\cc{FLOW001_0} (\autoref{fig:flow-0}) and \cc{FLOW001_1}
(\autoref{fig:flow-1}), are equivalent by synthesizing the following
relational invariant over their loop heads:
\[ \cc{T}_0 - \cc{x}_0 + \cc{T}_1 %
\]

\subsection{Results and conclusions}
\label{sec:exp-results}
\begin{center}
	\begin{table*}
		\centering
		\begin{tabular}{| r | r | r || r | r || r | r || r | r | r || r | r |  }
      \hline
      \multicolumn{3}{|c||}{\textbf{Benchmarks Features}}
      & \multicolumn{4}{c||}{ \textbf{\sys}}
      & \multicolumn{5}{c|}{\textbf{Baseline}}
      \\
      \hline
      \multicolumn{1}{|c|}{\textbf{Name}} &
      \multicolumn{1}{c|}{\textbf{Pairs}} &
      \multicolumn{1}{c||}{\textbf{LoC}} &
      \multicolumn{1}{c|}{\textbf{Eq.}} &
      \multicolumn{1}{c||}{\textbf{Time}} &
      \multicolumn{1}{c|}{\textbf{Ineq.}} &
      \multicolumn{1}{c||}{\textbf{Time}} &
      \multicolumn{1}{c|}{\textbf{TO}} &
      \multicolumn{1}{c|}{\textbf{Eq.}} &
      \multicolumn{1}{c||}{\textbf{Time}} &
      \multicolumn{1}{c|}{\textbf{Ineq.}} &
      \multicolumn{1}{c|}{\textbf{Time}} \\
      \hline
      \hline
      \cc{addDigits}
      & 1
      & 5
      & 1
      & 21.65s
      & 0
      & -
      & 0
      & 1
      & 13.23s
      & 0
      & -
      \\
      \hline
      \cc{ClimbStairs}
      & 3
      & 10
      & 3
      & 3m58s
      & 0
      & -
      & 3
      & 0
      & -
      & 0
      & -
      \\
      \hline
      \cc{ReverseInteger}
      & 1
      & 10
      & 1
      & 1m43s
      & 0
      & -
      & 1
      & 0
      & -
      & 0
      & -
      \\
      \hline
      \cc{trailingZero}
      & 4
      & 6.7
      & 4
      & 1m34s
      & 0
      & -
      & 4
      & 0
      & -
      & 0
      & -
      \\
      \hline
      \hline
      \cc{EX}
      & 1
      & 7
      & 1
      & 0.21s
      & 0
      & -
      & -
      & -
      & -
      & -
      & -
      \\
      \hline
      \cc{LWS}
      & 2
      & 57
      & 2
      & 1.81s
      & 0
      & -
      & -
      & -
      & -
      & -
      & -
      \\
      \hline
      \cc{DIVIDING}
      & 5
      & 24.6
      & 5
      & 4m1s
      & 0
      & -
      & -
      & -
      & -
      & -
      & -
      \\
      \hline
      \cc{ANUTHM}
      & 30
      & 30.3
      & 30
      & 2m40s
      & 0
      & -
      & -
      & -
      & -
      & -
      & -
      \\
      \hline
      \cc{AMIFIB}
      & 10
      & 34.8
      & 10
      & 28s
      & 0
      & -
      & -
      & -
      & -
      & -
      & -
      \\
      \hline
      \cc{FLOW002}
      & 58
      & 19
      & 51
      & 2m14s
      & 7
      & 3.32s
      & -
      & -
      & -
      & -
      & -
      \\
      \hline
      \cc{FLOW001}
      & 51
      & 19
      & 51
      & 2m1s
      & 0
      & -
      & -
      & -
      & -
      & -
      & -
      \\
      \hline
      \cc{START01}
      & 59
      & 11.6
      & 51
      & 0.26s
      & 8
      & 0.04s
      & -
      & -
      & -
      & -
      & -
      \\
      \hline
      \cc{MUFFINS3}
      & 61
      & 19.3
      & 51
      & 2m24s
      & 10
      & 2.54s
      & -
      & -
      & -
      & -
      & -
      \\
      \hline
      \cc{CIELAB}
      & 83
      & 24.5
      & 51
      & 22.13s
      & 32
      & 5.54s
      & -
      & -
      & -
      & -
      & -
      \\
      \hline
		 \end{tabular}
     \caption{The results of our evaluation of \sys.
       "Benchmarks Features" contains features of the
       subject program pairs, in particular the name of the problem that 
       the programs solve ("Name"), the numbers of pairs of
       solutions ("Pairs") checked, and the average lines of code
       ("LoC") in each solution.
       "\textsc{Pequod}" contains features of the
       performance of \sys, in particular the number of pairs of
       solutions check ("Eq."), the average time taken to prove equivalence
       ("Time"), the number of pairs of solutions proved,
       inequivalent ("Ineq.") and the average time of proving
       inequivalence ("Time"). 
       "\textsc{BaseLine}" contains features of the performance of \baseLine,
       in particular the number of pairs of solutions that timed out
       (the timeout limit was 500s.),
       the number of pairs of solutions proved
       equivalent ("Eq."), and the average time of proving equivalent
       ("Time"), over only pairs that did \emph{not} timeout.
       the number of pairs of solutions proved
       inequivalent ("Ineq."), and the average time of proving inequivalent
       ("Time"), over only pairs that did \emph{not} timeout.
      }
    \label{table:bench-large}
  \end{table*} 
\end{center}

We ran \sys to determine partial equivalence of the $369$ program
pairs collected.
We also ran \baseLine to determine partial equivalence
of the nine program pairs collected that did not operate on input and
output streams.
Because the rest of program pairs we collected has stream I/O
that hard to express the assertion in self-composition technical.
The results are contained in \autoref{table:bench-large}.
In \autoref{table:bench-large}, the first four problems are hosted on
Leetcode~\cite{leetcode} and the rest of the problems are hosted on
CodeChef~\cite{codechef}.

The only pair of programs that \baseLine can prove equivalent is the
pair of solutions to the Add Digits problem.
Both solutions to this problem have an input output relation that can
be described precisely by a formula in linear arithmetic.
\baseLine is able to infer such a formula automatically.
\sys requires more time to infer such a solution for the solutions to
Add Digits.
However, the additional time required by \sys to prove equivalence of
a relatively simple pair of programs seems to be an acceptable cost to
pay in many contexts in order to obtain the added power of \sys for
proving equivalence of more complex pairs of programs.

In summary, our results indicate that \sys significantly improves the
state of the art in verifying equivalence of concise, but subtle
alternative implementations.


\section{Related Work}
\label{sec:related-work}
Verifying the equivalence of two programs can also be reduced to
synthesizing and proving the correctness of a suitable \emph{product
  program}~\cite{barthe11,barthe16}.
Previous approaches construct the product program depending partly on
matching control structures between the pairs of programs and
establishing the logical equivalence of program conditions of matched
structures.
Previous work has also explored constructing \emph{asymmetric product
  programs}~\cite{barthe13} which can express proofs of equivalence
between programs with loops.
Such work does not address the problem of automatically inferring
loop invariants of the synthesized product program, which may be
viewed alternatively as relational invariants between loops of the
original programs.
This problem is directly addressed by \sys.

For programs $\cc{P}_0$ and $\cc{P}_1$, a special instance of the
product programs of $\cc{P}_0$ and $\cc{P}_1$ is the \emph{sequential
  composition} of $\cc{P}_0$ and $\cc{P}_1$.
Previous work has explored reducing verifying equivalence to
constructing the self-composition of given programs and proving that
it satisfies a suitable derived safety
property~\cite{barthe04,lopes16,terauchi05} or synthesizing sequential
summaries of the program by reduction to solving a system of
\emph{constrained Horn clauses (CHCs)}~\cite{felsing14}.
A key limitation of such approaches is that they can only infer proofs
of correctness that can be expressed using summaries of each program's
behavior in logic used by the verifier.
Such logics typically are not sufficiently strong to express summaries
required to prove the equivalence of non-trivial
programs~\cite{barthe11}, including the solutions to programming
problems that we encountered on online coding
platforms~\cite{codechef,leetcode}.
\sys attempts to synthesize relational invariants over internal
control locations of two programs.
Such a strategy enables \sys to prove partial equivalence of a larger
class of pairs of programs, both in principle (as discussed in
\autoref{sec:self-composition}) and in practice (as discussed in
\autoref{sec:exp-results}).

Previous work has proposed automatic verifiers of concurrent
programs~\cite{gupta11} that synthesize relational invariants by
generating a CHC system that is discharged with a generic CHC
solver~\cite{bjorner13,rummer13}.
\sys is similar to such approaches in that it attempts to construct a
proof of correctness from relational invariants over pairs of paths.
\sys is distinct from such approaches in that it uses a novel
construction of relational invariants that can be used to prove
partial equivalence of paths of independent programs (given in
\autoref{sec:find-path-invs}), and uses a novel algorithm that
constructs pairs of relational invariants over locations based on
relational invariants for pairs of paths (given in
\autoref{sec:solve}).

Several automatic equivalence checkers have been proposed for
verifying the equivalence of affine~\cite{verdoolaege11} and numerical
programs~\cite{partush13}.
\sys can be applied to programs that use any language features that
can be axiomatized in a logical theory with interpolation, such as
objects and arrays.
\sys does not require widening operations carefully tuned to
particular numerical domains in order to converge.

Several proof systems have been proposed in both
foundational~\cite{hoare69} and modern
work~\cite{hawblitzel13,sousa16} for proving total program
equivalence, simulation, and $k$-safety.
For given programs $\cc{P}_0$ and $\cc{P}_1$, such systems can express
proofs of equivalence by establishing the validity of semantic
summaries that relate the behavior of functions in $\cc{P}_0$ and
$\cc{P}_1$.
Regression-verification techniques~\cite{godlin09} match substructures
of a pair of programs based on a traversal of the programs' syntactic
structure and attempt to prove that matched substructures are
equivalent, using provided candidate relational invariants.
Regression verification can be optimized, using symbolic execution to
only analyze slices of two given versions of a program that are
changed~\cite{backes13}.
Regression verification can also be applied to partitions of the given
programs' input space, defined by path formulas of individual program
paths, enabling programs to be proved equivalent
gradually~\cite{bohme13}.

Recent work has provided logic systems for reasoning about relational
properties of higher-order programs~\cite{aguirre17}.
However, these systems have not yet been used to automatically
synthesize proofs of program equivalence.
\sys can only infer proofs in a space of structures that is less
expressive than the proof structures proposed in such work: in
particular, the proofs inferred by \sys are evidence of only
\emph{partial} equivalence.
However, \sys attempts to synthesize such proofs automatically.

Several approaches have been proposed that attempt to verify the
equivalence of programs $\cc{P}_0$ and $\cc{P}_1$ by symbolically
executing the paths of $\cc{P}_0$ and $\cc{P}_1$.
\textsc{SymDiff} verifies that given programs that are
loop-free~\cite{lahiri12} or that are annotated with synchronization
points~\cite{lahiri13} satisfy expected relational summaries.
Unlike \textsc{SymDiff}, \sys may not always terminate, but \sys can
be applied to potentially prove the partial equivalence of programs
with loops.
\textsc{UCKlee}, similar to \sys, symbolically executes both programs
and inspects pairs of path formulas for control paths of $\cc{P}_0$
and $\cc{P}_1$ to determine if they are paths on which $\cc{P}_0$ and
$\cc{P}_1$ are not equivalent~\cite{ramos11}.
However, \sys can also potentially use the proofs of equivalence of a
pair of paths to prove that given programs are equivalent.

A \emph{differential symbolic execution} engine~\cite{person08}
symbolically executes given programs $\cc{P}_0$ and $\cc{P}_1$, and can
optionally construct a formula for each program that over-approximates
the effect of each.
The engine then compares the relational formulas for each program to
determine if the programs may be equivalent.
Such an engine is similar to \sys, in that it uses symbolic reasoning
to attempt to automatically synthesize a sound over-approximation of
the effect of each program.
However, a key distinction between such an engine and \sys is that
\sys infers relational invariants between programs by iteratively
selecting and analyzing particular paths, rather than computing a
fixed over-approximation of each program and then comparing the
approximations.

Analyses for \emph{rootcausing} failures of program
equivalence~\cite{lahiri15} take a pair of control paths that prove
the non-equivalence of two programs and generate a minimal-cost change
to the programs that removes the feasibility of the counterexample.
Similarly to rootcausing analyses, \sys applies a precise symbolic
analysis to pairs of control paths from $\cc{P}_0$ and $\cc{P}_1$.
Unlike rootcausing analyses, \sys analyzes control paths either to
determine that the paths are a true counterexample to equivalence or
to synthesize path invariants that prove that the control paths are
equivalent.

Several techniques have been proposed that improve the effectiveness
of static program analyses by analyzing multiple versions of a
program.
The \emph{differential-assertion-checking} problem~\cite{lahiri13} is
to determine if one version of a given program satisfies all
assertions satisfied by a previous version of the program.
\emph{Verification modulo versions}~\cite{logozzo14} filters warnings
generated by applying a static analysis to a new version of a program
to only the warnings that are novel to the new version.
Optimizations to static analysis have been proposed that compute
function summaries using an interpolating theorem
prover~\cite{sery12};
when analyzing a new version of the program, the optimized analysis
first checks if the summaries computed for functions in the original
version of the program are valid summaries for functions in the new
version of the program.
All of the above approaches use multiple versions of a program to
optimize the behavior of a safety analysis;
these problems are distinct from the problem addressed by \sys, which
is to determine if two programs are partially equivalent.
In particular, while \sys also synthesizes an abstraction of given
programs from interpolants, the interpolants are synthesized from
proofs that pairs of paths from \emph{multiple} programs are partially
equivalent.

Some software model checkers select a program abstraction by
constructing Craig
interpolants~\cite{albarghouthi12,heizmann10,mcmillan04,mcmillan06,rummer13}
of sub-formulas of formulas that characterize runs of individual
paths.
However, unlike the above techniques \sys uses interpolants to prove
the equivalence of paths selected from \emph{distinct} programs.

Previous work has identified equivalence verification as a problem
with critical applications in programming education, and has proposed
\emph{autograding} techniques for automatically editing a student
solution so that it is equivalent to a reference
solution~\cite{singh13}.
Existing work on autograding relies on a bounded model checker to
determine if programs may be equivalent.
An autograder that uses an improved equivalence verifier would enjoy a
stronger soundness guarantee for determining when a student's solution
is correct.
An autograder designed to use not just counterexamples to equivalence
but also relational invariants for equivalence could potentially
suggest edits to student solutions that are functionally correct but
could be simplified or optimized.


\section{Conclusion}
\label{sec:conclusion}
We have presented a novel algorithm that attempts to prove the partial
equivalence of given programs.
A key challenge in proving the partial equivalence of given programs
$P_0$ and $P_1$ is to both synthesize a suitable product program $P_1$
of $P_0$ and $P_1$, and to synthesize inductive invariants of $P'$
that prove the equivalence of $P_0$ and $P_1$.
Previous approaches address this problem by first choosing a product
program either by choosing one from a heavily restricted class of
product programs, requiring a product program to be given manually, or
choosing one based on fixed heuristics.
After choosing a candidate product program, such approaches then
attempt to synthesize its inductive invariants.

We have presented a novel equivalence verifier, named \sys, that does
not operate under any of the above limitations.
The key feature of \sys is that it attempts to synthesize a product
program and its invariants simultaneously.
To do so, \sys iteratively collects proofs of equivalence of pairs of
paths of given programs, and attempts to extract a product program and
its inductive invariants from the invariants defined per pair of
paths.
We have implemented a prototype version of \sys that targets JVM
bytecode, and used it to verify hundreds of alternate solutions
submitted by students to online coding problems.


\bibliographystyle{ACM-Reference-Format}
\small
\setlength{\bibsep}{3pt}
\bibliography{p,conf}

\appendix

\section{Proof Of Correctness}
\label{app:proofs}
Here, we give a formal proof for \autoref{thm:soundness} and its
lemmas. First we provide the proof of the underlying lemmas, then we
provide the proof of the theorem.

The following is a proof of \autoref{lemma:valid-proof}:

\begin{proof}
  If there exists $I_0, I_1 \in \locinvs{ \cc{P}_0 }{ \cc{P}_1 }$, then
  the definition of location-pair
  invariants (\autoref{defn:location-pair-invs}) implies that for each pair of
  runs of $\cc{P}_0$ and $\cc{P}_1$ under the same input,
  $\cc{P}_0$ and $\cc{P}_0$ do not produce different output.
  The definition of path (\autoref{defn:path}), of run
  (\autoref{defn:runs}), and of partial equivalence
  (\autoref{defn:part-equiv}) prove that $\cc{P}_0 \equiv
  \cc{P}_1$.
\end{proof}

The following is a proof of \autoref{lemma:find-inv-corr}:

\begin{proof}
  There are two cases for this proof.

  For all $p_0 \in \paths{ \cc{P}_0 }$ and $p_1 \in \paths{ \cc{P}_1
  }$, if $p_0 \equiv p_1$, then $\refine(\cc{P}_0, \cc{P}_1, p_0,
  p_1) \in \pathpairinvs{p_0}{p_1}$.
  If $p_0 \equiv p_1$, then by the definition of partial
  equivalence (\autoref{defn:part-equiv}), for all pairs of paths
  $p_0 \in \paths{ \cc{P}_0 }$ and $p_1 \in \paths{ \cc{P}_1}$,
  $\noneqpaths{p_0}{p_1}$ is not satisfiable.
  By the core algorithm of \sys(\autoref{alg:sys}, \autoref{line:ret-equiv}), 
  $\refine(\cc{P}_0, \cc{P}_1, p_0,  p_1) \in \pathpairinvs{p_0}{p_1}$.

  For all $p_0 \in \paths{ \cc{P}_0 }$ and $p_1 \in \paths{ \cc{P}_1
  }$, if $p_0 \not \equiv p_1$, then $\pathpairinvs{p_0}{p_1} = \nonequiv$.
  If $p_0 \not \equiv p_1$, then by the definition of partial
  equivalence (\autoref{defn:part-equiv}), there exists a pair of
  paths $p_0 \in \paths{ \cc{P}_0 }$ and $p_1 \in \paths{
  \cc{P}_1}$, such that $\noneqpaths{p_0}{p_1}$ is satisfiable.
  By the core algorithm of \sys(\autoref{alg:sys}, \autoref{line:ret-inequiv}), 
  $\pathpairinvs{p_0}{p_1} = \nonequiv$.
\end{proof}

The following is a proof of \autoref{lemma:find-ind-corr}:

\begin{proof}
  We construct this proof by induction on the evaluation of
  $\chkinductive$ run over $\obligations$ and $\discharged$:

  The inductive claim is that if the path-pair invariants in
  $\obligations$ are inductive, then there exists a restriction on $I$
  which is a set of inductive path-pair invariants that contains all
  elements of $\discharged$.

  For the base case, $\chkinductive$ is called on
  $([ \initloc ], [ \initloc ])$ and
  $\emptyset$ (\autoref{alg:chk-ind}), which combined with the
  definition of inductive path-pair invariants
  (\autoref{defn:path-pair-invs}), implies the claim.
  %

  For the inductive case, when $\obligations$ is non-empty, a
  path-pair invariant $I(p_0, p_1) \in \obligations$ is inspected.
  $\obligations'$ is constructed by removing $I(p_0, p_1)$ from
  $\obligations$, and $\discharged'$ is constructed by adding $I(p_0,
    p_1)$ to $\discharged$. From here, there are two possibilities:

  If $I(p_0, p_1)$ is entailed by $I(p_0', p_1')$, $p_0$ and $p_0'$
  end with the same control location, $p_1$ and $p_1'$ end with
  the same control location, and $I(p_0', p_1') \in \discharged$
  then $\chkinductive$ calls itself recursively with
  $\obligations'$ and $\discharged'$ (\autoref{alg:chk-ind}, \autoref{line:chkind-rec}).
  This step maintains the inductive claim.
  Location-pair invariants are constructed by taking the
  disjunction of all path-pair invariants that end with the same
  control location. This fact, the fact that $p_0'$ and $p_1$ end
  with the same control location as $p_0$ and $p_1$, and the fact
  $I(p_0', p_1')$ entails $I(p_0, p_1)$ together indicate that
  $I(p_0, p_1) \lor I(p_0', p_1')$ still entails $I(p_0, p_1)$.
  Because $I(p_0, p_1) \lor I(p_0', p_1')$ hold for all clauses in
  the location-pair invariant system, the claim is established by
  definition of inductive path-pair invariants.

  Otherwise, $\chkinductive$ calls itself recursively on
  $\obligations'$ extended with the path-pair invariant from
  taking a step in the left program (\autoref{alg:chk-ind},
  \autoref{line:chkind-rec-0}) or in the right program
  (\autoref{alg:chk-ind}, \autoref{line:chkind-rec-1}) together
  with $\discharged'$.
  In these cases, the claim is established by the definition of
  inductive invariants (\autoref{defn:inductive-path-pair-invers})
  and the definition of location-pair invariants rules 2 and 3
  respectively (\autoref{defn:location-pair-invs}).

  When $\obligations$ is empty, $\chkinductive$ returns $\isind$, by
  \autoref{alg:chk-ind}.
  This fact, together with the inductive claim, implies
  that $I$ is a set of inductive path-pair invariants for $\cc{P}_0$
  and $\cc{P}_1$.
\end{proof}

As stated by \autoref{thm:soundness}, whenever \sys returns a
definite result, the result is correct.

\begin{proof}
  First we prove:
  For all $\cc{P}_0, \cc{P}_1 \in \lang$,
  if $\sys(\cc{P}_0, \cc{P}_1)$ is defined and $\cc{P}_0 \equiv
  \cc{P}_1$, then $\sys(\cc{P}_0, \cc{P}_1) = \true$.
  This can be restated as:
  for all $\cc{P}_0, \cc{P}_1 \in \lang$, 
  if $\sys(\cc{P}_0, \cc{P}_1)$ is defined and $\sys(\cc{P}_0, \cc{P}_1) = \false$,
  then $\cc{P}_0 \not \equiv \cc{P}_1$.
  If $\sys(\cc{P}_0, \cc{P}_1) = \false$, then the core algorithm \sys(\autoref{alg:sys}, \autoref{line:ret-inequiv}),
  implies that there exists $p_0 \in \paths{ \cc{P}_0 }$ and
  $p_1 \in \paths{ \cc{P}_1}$ such that $\pathpairinvs{p_0}{p_1} = \nonequiv$.
  Therefore, by \autoref{lemma:find-inv-corr}, $\cc{P}_0 \not \equiv \cc{P}_1$.

  Next we prove:
  For all $\cc{P}_0, \cc{P}_1 \in \lang$, %
  if $\sys(\cc{P}_0, \cc{P}_1)$ is defined and $\sys(\cc{P}_0, \cc{P}_1) = \true$, then
  $\cc{P}_0 \equiv \cc{P}_1$.
  If $\sys(\cc{P}_0, \cc{P}_1) = \true$, then the core algorithm \sys(\autoref{alg:sys}, \autoref{line:ret-equiv}),
  implies that $\chkinductive(\cc{P}_0, \cc{P}_1, I) = \isind$.
  By \autoref{lemma:find-ind-corr}, there exists some
  restriction of $I$ which is a set of inductive path-pair invariants of $\cc{P}_0$
  and $\cc{P}_1$.
  By the definition of inductive path-pair invariants
  (\autoref{defn:inductive-path-pair-invers}) and of location-pair
  invariants (\autoref{defn:location-pair-invs}), there exists $I_0,
  I_1 \in \locinvs{ \cc{P}_0 }{ \cc{P}_1 }$.
  Therefore, by \autoref{lemma:valid-proof}, $\cc{P}_0 \equiv \cc{P}_1$.
\end{proof}

\end{document}